\documentclass{aa}  

\usepackage{graphicx}
\usepackage{txfonts}
\usepackage{bbm}
\usepackage{url}
\usepackage{natbib}
\usepackage[breaklinks=true]{hyperref} 
\hypersetup{colorlinks=true,urlcolor=blue,citecolor=blue,pdfborder= 0 0 0}
\bibpunct{(}{)}{;}{a}{}{,}    
\begin{document}

   \title{TOPz: Photometric redshifts for J-PAS}


   \author{J.~Laur\inst{1}  \and
          E.~Tempel\inst{1,2} \and
          A.~Tamm\inst{1} \and
          R.~Kipper\inst{1} \and
          L.~J.~Liivam\"agi\inst{1} \and
          A. Hern\'an-Caballero\inst{3} \and
          M.~M.~Muru\inst{1} \and
          J. Chaves-Montero\inst{4} \and
          L.~A.~D\'iaz-Garc\'ia\inst{5} \and
          S.~Turner\inst{1} \and
          T.~Tuvikene\inst{1} \and
          C. Queiroz\inst{6}\and
          C. R. Bom\inst{7,8} \and
          J. A. Fern\'andez-Ontiveros\inst{3} \and
          R.~M.~Gonz\'alez Delgado\inst{5} \and
          T. Civera\inst{3} \and
          R. Abramo\inst{9} \and
          J. Alcaniz\inst{10} \and
          N. Benitez\inst{5} \and
          S. Bonoli\inst{3,4} \and
          S. Carneiro\inst{11} \and
          J. Cenarro\inst{12} \and
          D. Crist\'obal-Hornillos\inst{12} \and
          R. Dupke\inst{10} \and
          A. Ederoclite\inst{3}\and
          C. L\'opez-Sanjuan\inst{12} \and
          A. Mar\'in-Franch\inst{3} \and
          C. M. de Oliveira\inst{9} \and
          M. Moles\inst{3} \and
          L. Sodr\'e Jr.\inst{9} \and
          K. Taylor\inst{13} \and
          J. Varela\inst{12} \and
          H. V. Rami\'o\inst{12}
          }

   \institute{
   Tartu Observatory, University of Tartu, Observatooriumi 1, 61602 T\~oravere, Estonia\\
              \email{jaan.laur@ut.ee}
			  \and
			  Estonian Academy of Sciences, Kohtu 6, 10130 Tallinn, Estonia
			  \and
			  Centro de Estudios de F\'isica del Cosmos de Arag\'on (CEFCA), Plaza San Juan, 1, E-44001 Teruel, Spain
			  \and
			  Donostia International Physics Center, Paseo Manuel de Lardiz\'abal 4, E-20018 Donostia-San Sebastián, Spain
			  \and
			  Instituto de Astrof\'{\i}sica de Andaluc\'{\i}a (CSIC), P.O.~Box 3004, 18080 Granada, Spain
              \and
              Departamento de Astronomia, Instituto de F\'isica, Universidade Federal do Rio Grande do Sul (UFRGS), Av. Bento Gon\c{c}alves, 9500, Porto Alegre, RS, Brazil
              \and
              Centro Brasileiro de Pesquisas F\'isicas, Rua Dr. Xavier Sigaud 150, CEP 22290-180, Rio de Janeiro, RJ, Brazil
              \and
              Centro Federal de Educa\c{c}\~ao Tecnol\'ogica Celso Suckow da Fonseca, Rodovia M\'ario Covas, lote J2, quadra J, CEP 23810-000,  Itagua\'i, RJ, Brazil
              \and
              Instituto de F\'isica, Universidade de S\~aoo Paulo, Rua do Mat\~aoo 1371, 05508-090 S\~aoo Paulo, Brazil
              \and
              Observat\'orio Nacional, Minist\'erio da Ciencia, Tecnologia, Inova\c{c}\~ao e Comunica\c{c}\~oes, Rua General Jos\'e Cristino, 77, S\~ao Crist\'ov\~ao, 20921-400, Rio de Janeiro, Brazil
              \and
              Instituto de F\'isica, Universidade Federal da Bahia, 40210-340, Salvador, BA, Brazil
              \and
              Centro de Estudios de de F\'isica del Cosmos de Arag\'on (CEFCA), Unidad Asociada al CSIC, Plaza San Juan, 1, 44001 Teruel, Spain
              \and
              Instruments4, 4121 Pembury Place, La Canada Flintridge, CA 91011, U.S.A.
             }


 
  \abstract
   {The importance of photometric galaxy redshift estimation is rapidly increasing with the development of specialised powerful observational facilities.}
   {We develop a new photometric redshift estimation workflow TOPz to provide reliable and efficient redshift estimations for the upcoming large-scale survey J-PAS which will observe 8500 deg$^2$ of the northern sky through 54 narrow-band filters.} 
   {TOPz relies on template-based photo-$z$ estimation with some added J-PAS specific features and possibilities.
   We present TOPz performance on data from the miniJPAS survey, a precursor to the J-PAS survey with an identical filter system. 
   First, we generated spectral templates based on the miniJPAS sources using the synthetic galaxy spectrum generation software \textit{CIGALE}.
   Then we applied corrections to the input photometry by minimising systematic offsets from the template flux in each filter. 
   To assess the accuracy of the redshift estimation, we used spectroscopic redshifts from the DEEP2, DEEP3, and SDSS surveys, available for 1989 miniJPAS galaxies with $r<22$ $\mathrm{mag_{AB}}$. 
   We also tested how the choice and number of input templates, photo-$z$ priors, and photometric corrections affect the TOPz redshift accuracy.}
   {The general performance of the combination of miniJPAS data and the TOPz workflow fulfills the expectations for J-PAS redshift accuracy. 
   Similarly to previous estimates, we find that 38.6\% of galaxies with $r<22$ mag reach the J-PAS redshift accuracy goal of d$z/(1 + z) < 0.003$. 
   Limiting the number of spectra in the template set improves the redshift accuracy up to 5\%, especially for fainter, noise-dominated sources.
   Further improvements will be possible once the actual J-PAS data become available.}
   {}


   \maketitle
%
\section{Introduction}

Photometric galaxy redshift surveys provide a viable alternative and a complement to spectroscopic surveys for acquiring a 3D map of the Universe and they play an important role in studies of galaxy evolution, galaxy clusters, stellar populations, and star formation rates.
While the precision of a photometrically estimated redshift (photo-$z$) typically lags behind that of a spectroscopically determined one, the considerable gain in speed and the lack of source selection effects favour photo-$z$ surveys for cosmological applications where large, deep, and unbiased data sets are desired.
Due to these observational advantages, photo-$z$s will be the most viable solution for redshift estimations in the imminent big data era of galaxy and cosmological surveys.
Recent years have brought along a multitude of new observational initiatives and instrumentation for obtaining multi-band photometry of large parts of the sky: ALHAMBRA \citep{2008AJ....136.1325M}, PAU \citep{2009ApJ...691..241B}, J-PAS \citep{2014arXiv1403.5237B}, HSC-SSP \citep{2018PASJ...70S...4A}, J-PLUS \citep{2019A&A...622A.176C}, S-PLUS \citep{2019MNRAS.489..241M, 2022MNRAS.511.4590A}, and KiDS \citep{2020A&A...633A..69H}.

While the derivation of redshifts from spectroscopic data is relatively straightforward, the situation is quite different for photometric redshifts (photo-z), where spectral features may easily remain undetected, unresolved, or misidentified. Recent years have seen a big leap forward in overcoming these obstacles and a large variety of photo-$z$ estimation methods and algorithms have been developed; we refer readers to \citet{2019NatAs...3..212S} for a recent overview.

Most broadly, photo-$z$ methods can be split between those based on machine-learning \citep[e.g.][]{2013MNRAS.432.1483C,2014MNRAS.442.3380C,2015MNRAS.449.2040H,2016PASP..128j4502S,2018MNRAS.475..331G,2018AJ....155....1G} and those based on spectral templates \citep[e.g.][]{2008ApJ...686.1503B,2014MNRAS.441.2891M,2017A&C....19...34B,2000ApJ...536..571B,2021A&A...650A..90A}. 
Performance-wise, no clear winner has yet emerged, mostly due to various assumptions underlying each estimation approach \citep{2020MNRAS.499.1587S}. 

Theoretically, machine-learning algorithms are capable of using all the information available in the data and should thus yield maximal possible accuracy. 
In addition, machine-learning algorithms tend to be faster than template-based ones.
However, their performance generally depends on the size and quality of the training set, which becomes problematic at higher redshifts, where an unbiased comprehensive observational data set is hard to obtain; thus machine-learning algorithms are generally outperformed by template-based methods in this regime \citep{2010A&A...523A..31H}. 
In addition, template-based methods have another advantage in that they may simultaneously be used to derive a range of physical properties of galaxies via spectral energy distribution (SED) fitting \citep{2011Ap&SS.331....1W, 2015A&A...582A..14D, 2019ApJ...882...61B, 2019A&A...631A.156D, 2021A&A...649A..79G}.

Most typically, photo-$z$ methods have been applied to broad-band filter data \citep[][and many others]{2006A&A...457..841I, 2018PASJ...70S...9T, 2020MNRAS.497.1935L}. 
In such cases, redshifts are primarily derived from large-scale features such as the $4000$~\AA, Balmer, or Lyman breaks if they are present in between the filter passband ranges.  
Thereby, a redshift accuracy of up to d$z$ $\approx0.03(1 + z)$ has been achieved. 
Using narrow-band filters, this limit can be considerably extended by a more precise localisation of the continuum breaks as well as by the possibility of detecting individual spectral lines: JPLUS/SPLUS \citep{2019A&A...631A..82I}, miniJPAS \citep{2021A&A...653A..31B}, and PAU \citep{2020MNRAS.497.4565E, 2020A&A...634A.123R}.

In essence, the template-based photo-$z$ methods share the general working principle of measuring the $\chi^2$ deviation between the observed SED and each template on a redshift grid for a given source. 
However, in practical applications, the performance may be strongly influenced by the library of the spectral templates \citep{2015MNRAS.451.1848G, 2020A&A...637A.100W}, the applied redshift prior \citep[in the case of Bayesian methods, see][]{2000ApJ...536..571B, 2014MNRAS.441.2891M, 2015ApJ...801...20T}, or the method by which the best redshift estimate is extracted. In particular, to make the most of narrow-band filter observations of high redshift sources, the spectral templates should also cover the UV part of the spectrum.
These aspects favour synthetic spectra over observational ones when it comes to templates.

We present a new template-based photometric redshift workflow TOPz (Tartu Observatory photo-$z$), which was purpose-built for the J-PAS survey, but easily applicable to any other galaxy photometry data set.
The main goal is to include improvements to photo-z estimation that are specific to J-PAS observations along with a cluster-ready implementation to speedily handle the expected amount of data.
We provide an overview of the method, a recipe for generating a suitable set of template spectra, and the results of an application on the current miniJPAS data set.

The outline of the paper is as follows. 
In Sect.~\ref{sec:bayesian_big} we give a brief overview of the Bayesian photometric redshift estimation method and in Sect.~\ref{sec:topz} an overview of our photo-$z$ workflow TOPz.
In Sect.~\ref{sec:mini_jpas} we describe the miniJPAS data. 
The construction of the templates, photometric corrections, and photo-$z$ priors are described in Sect.~\ref{sec:templates}.
The impact of the aforementioned inputs along with the results are given in Sect.~\ref{sec:minijpas_z} and a discussion follows in Sect.~\ref{sec:conclusion}.

\section{Bayesian photometric redshift estimation}
\label{sec:bayesian_big}
\subsection{General overview}
\label{sec:bayesian}

In a Bayesian framework, the problem of photo-$z$ estimation can be posed as finding the probability of a galaxy having redshift $z$ given the observational data and some prior information \citep{2000ApJ...536..571B}. 
The probability can be expressed as
\begin{equation}
    p(z\,|\,D,I),
    \label{eq:pz_single}
\end{equation}
where $D=\{F,m_0\}$ describes the observed SED that is given by relative fluxes $F$ at different wavelengths and total magnitude $m_0$ in a reference passband.
The latter also sets the absolute scale for the SED.
The term $I$ includes the prior information not already contained in $D$.
To simplify the mathematical notation, we drop the term $I$ below.

Expression \eqref{eq:pz_single} gives the probability assuming that a galaxy has a known spectral type.
The measured SED $F$ of a galaxy can be approximated with a variety of different spectral types, represented by a set of spectral templates $\mathbf{T}$. 
A given galaxy cannot belong to two spectral types at the same time, thus the probability in expression \eqref{eq:pz_single} can be expanded as $p(z,T\,|\,D)$, that is the probability of the galaxy redshift being $z$ while the galaxy has a type $T$. 
This can in turn expanded as
\begin{equation}
    p(z,T\,|\,D) =  p(z,T\,|\,F,m_0)\propto  p(z,T\,|\,m_0)p(F\,|\,z,T),
    \label{eq:pfinal}
\end{equation}
where in the second step we applied the Bayes' theorem. 
The last expression $p(F\,|\,z,T)$ gives the probability that the measured relative fluxes $F$ correspond to the template $T$ at redshift $z$. 
We assume that the probability does not depend on the magnitude $m_0$. 
The prior $p(z,T\,|\,m_0)$ can be further developed using the product rule
\begin{equation}
    p(z,T\,|\,m_0) = p(T\,|\,m_0)p(z\,|\,T,m_0),
    \label{eq:prior}
\end{equation}
where $p(T\,|\,m_0)$ is the general, independently known galaxy type fraction as a function of galaxy magnitude and $p(z\,|\,T,m_0)$ is the general redshift distribution of galaxies with spectral template $T$ and magnitude $m_0$.

The template-dependent redshift posterior for a given galaxy with total magnitude $m_0$ and observed fluxes $F$ is defined in Eq.~\eqref{eq:pfinal}. 
The calculation of $p(z,T\,|\,F,m_0)$ assumes that we know how to calculate the redshift likelihood $p(F\,|\,z,T)$ and how to estimate the prior $p(z,T\,|\,m_0)$. 
The former is explained in Sect.~\ref{sec:likelihood}, the latter is discussed in Sect.~\ref{sec:prior}.
To find the redshift posterior, we can marginalise over the template set $\mathbf{T}$
\begin{equation}
    p(z\,|\,F,m_0) = \sum\limits_{T\in \mathbf{T}} p(z,T\,|\,F,m_0).
    \label{eq:p}
\end{equation}

\subsection{Estimating the redshift likelihood}
\label{sec:likelihood}

Following \citet{2000ApJ...536..571B} the redshift likelihood of a galaxy $p(F\,|\,z,T)$ can be written as
\begin{equation}
    p(F\,|\,z,T) \propto \sqrt{F_\mathrm{TT}}\exp\left[ -\frac{\chi^2(z,T,a)}{2} \right],
    \label{eq:likelihood}
\end{equation}
where normalisation factor $F_\mathrm{TT}$ (see Eq.~\eqref{eq:FTT}) comes from the integration over nuisance parameter $a$. 
The $\chi^2$ defines the quantity to be minimised and is defined as
\begin{equation}
    \chi^2(z,T,a) = \sum\limits_{j=1}^{N_\mathrm{filt}} \frac{ \left(F_j-aF_{T,j}\right)^2 }{\sigma^2_{F_j}},
    \label{eq:chi2}
\end{equation}
where summation is over all observed passbands for a given galaxy.
$F_j$ and $\sigma_{F_j}$ are the observed galaxy flux and its standard deviation through passband $j$ while $F_{T,j}$ is the synthetic flux of a redshifted template $T$ through passband $j$.

Equation~\eqref{eq:chi2} can be rewritten as
\begin{equation}
    \chi^2(z,T,a) = F_\mathrm{OO} - \frac{F^2_\mathrm{OT}}{F_\mathrm{TT}}+(a-a_m)^2F_\mathrm{TT},
    \label{eq:chi2_ver2}
\end{equation}
where
\begin{equation}
    a_m = \frac{F_\mathrm{OT}}{F_\mathrm{TT}}
\end{equation}
is the value of nuisance parameter $a$ that minimises Eq.~\eqref{eq:chi2} and Eq.~\eqref{eq:chi2_ver2}.
The notations $F_\mathrm{OO}$, $F_\mathrm{OT}$, and $F_\mathrm{TT}$ are defined as
\begin{eqnarray}
    F_\mathrm{OO}&=&\sum_{j=1}^{N_\mathrm{filt}}\frac{F_j^2}{\sigma^2_{F_j}}, \\
    F_\mathrm{OT}&=&\sum_{j=1}^{N_\mathrm{filt}}\frac{F_jF_{T,j}}{\sigma^2_{F_j}}, \\
    F_\mathrm{TT}&=&\sum_{j=1}^{N_\mathrm{filt}}\frac{F_{T,j}^2}{\sigma^2_{F_j}}.
    \label{eq:FTT}
\end{eqnarray}

For a given galaxy with observed fluxes $F_j$, we calculate the likelihood (defined in Eq.~\eqref{eq:likelihood}) that the observational data correspond to a given template $T$ at a given redshift $z$.
For practical reasons, the redshift is mapped onto a user-defined grid and the template set $\mathbf{T}$ contains a limited number of templates.

\section{TOPz}
\label{sec:topz}

\subsection{Overview}
\label{sec:topz_overview}

\begin{figure}
    \centering
    \includegraphics[width=\linewidth]{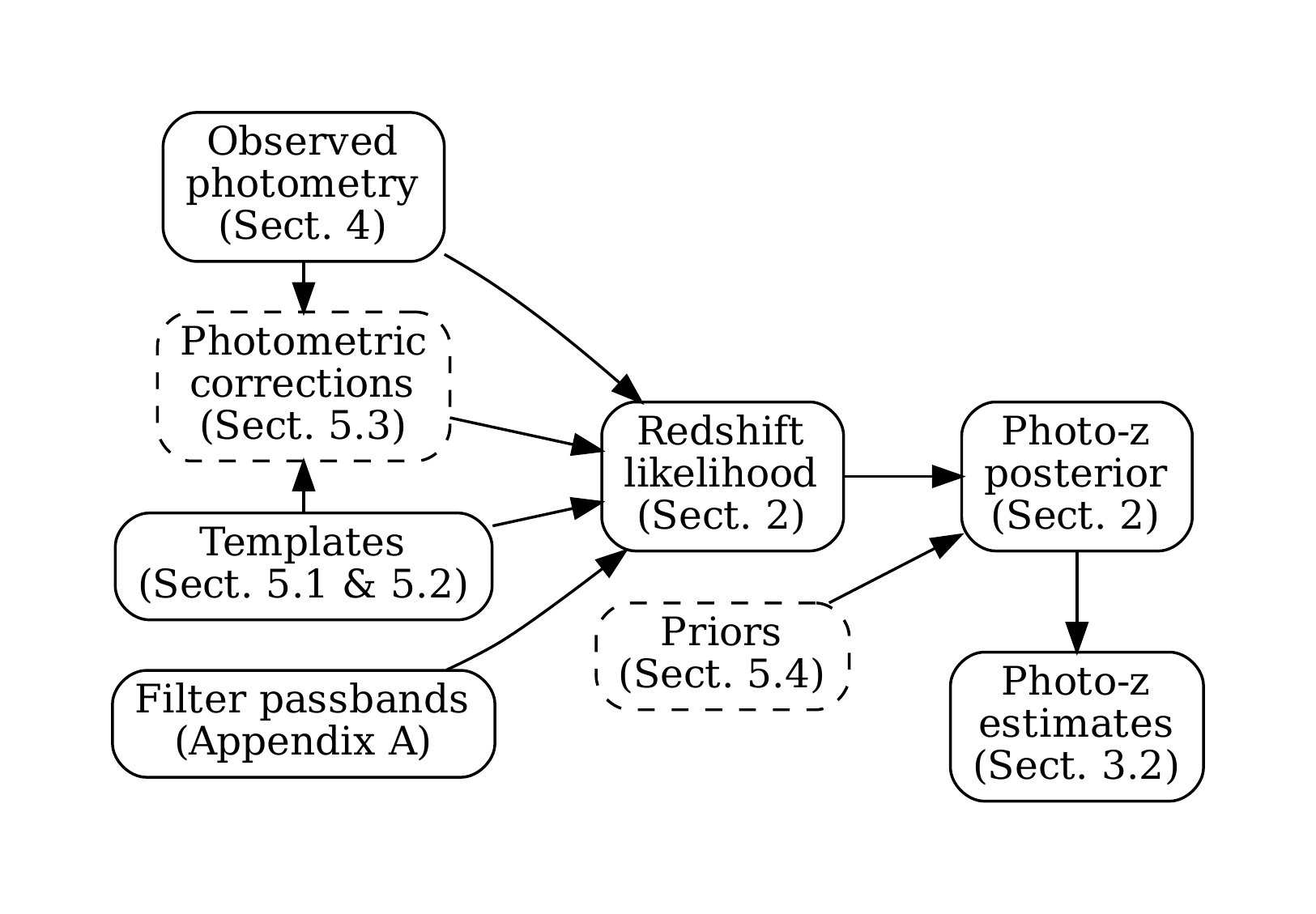} 
    \caption{Flowchart describing the basic functionality of a template based photo-$z$ code. The section numbers below each element refer to the outline of this paper. The dashed boxes indicate optional inputs.}
    \label{fig:flowchart}
\end{figure}

The basic workflow of TOPz along with references to the corresponding sections in this paper can be seen in Fig.~\ref{fig:flowchart}.
TOPz follows the Bayesian approach described in Sect.~\ref{sec:bayesian_big} by evaluating the likelihood of a galaxy lying at certain redshifts.
The likelihood is calculated based on a preselected set of templates that best describe the spectral types of observed galaxies in a given data set.
Synthetic photometry can be calculated by combining the observed optical system transmission curve (passband) with the spectral templates. 
This transmission curve usually incorporates the CCD quantum efficiency, filter transmission, atmospheric transmission, and the optics of a particular telescope.
For each galaxy, the templates are redshifted along a grid and the corresponding $\chi^2$ values (Eq.~\eqref{eq:chi2}) are used to find the redshift likelihood $p(F\,|\,z,T)$. 
This likelihood can be further refined by introducing a prior, for example on the basis of previously determined redshift distribution of galaxies.
From the likelihood distribution, different estimations for the `best' redshift can be inferred as described in Sect.~\ref{sec:benefits}.

In order to run TOPz for the redshift estimations, there are multiple required inputs (see Fig.~\ref{fig:flowchart}).
First input is a catalogue containing the fluxes of the observed galaxies in each passband along with the corresponding uncertainties (see Sect.~\ref{sec:mini_jpas}). 
TOPz does not require all the passbands for each galaxy to be present as specific passbands can be flagged and thus ignored in the fitting process.
Second inputs are the effective transmission curves for every filter used in the observations, ideally combining the filter transmission, CCD quantum efficiency, effects from telescope optics and atmospheric transmission.
Third input it the set of spectral templates (see Sect.~\ref{sec:templates}).

As an optional input, a set of photometric corrections can be provided to be applied to the input catalogue (see Sect.~\ref{sec:photo_corr}). 
This is introduced to correct systematic errors in the observations.
Photometric corrections can be given separately for each galaxy or as a single value for the whole data set.
Systematic uncertainties for each filter can also be adjusted in this phase.
Another optional inputs are the priors for modifying the redshift likelihood depending on the brightness and type of the observed galaxy (see Sect.~\ref{sec:prior}). 
Separate priors can be used for each spectral template or groups of templates.

TOPz follows the general Bayesian likelihood calculation workflow of the BPZ code\footnote{https://www.stsci.edu/~dcoe/BPZ/} while also using the same priors defined in \citet{2000ApJ...536..571B}. 
However, TOPz differs by its ability to generate and prioritise templates based on the observed photometry, allowing to reach a better correspondence between the template set and the actual occurrence statistics and photometry of the sources. In addition, these templates can also be used to calibrate the observed photometry and thereby further improve the resultant photo-z accuracy.

One of the unique features of TOPz is a J-PAS specific option to consider multiple passbands per filter. 
This option enables to take into account the dependency of filter transmission curves on the incident angle of the light, arising in the J-PAS optical system due to large field of view \citep{2014arXiv1403.5237B}.
When looking at an observation through a single filter in J-PAS, each galaxy will have a different passband that will be constructed based on the galaxy's position on the frame as well as on the information how that specific tile has been observed. 
For a more in-depth analysis on the impact of this effect see Appendix~\ref{sec:CW}.

TOPz is mainly written in Fortran language and has been developed keeping in mind the forthcoming J-PAS data set, expected to contain SED measurements for around $10^7$ galaxies which is $\sim17500$ galaxies per $deg^2$ \citep{2021A&A...654A.101H}.
Therefore, the code is built with parallel computing capabilities and designed to run on a computer cluster.
We evaluated that, using 100 spectral templates on a 30-core machine, it would take around three minutes to calculate photo-$z$ estimates for 1 $deg^2$ and around 30 hours to estimate them for the whole catalogue of $10^7$ galaxies.

\subsection{Outputs}
\label{sec:benefits}

\begin{figure}
    \centering
    \includegraphics[width=\linewidth]{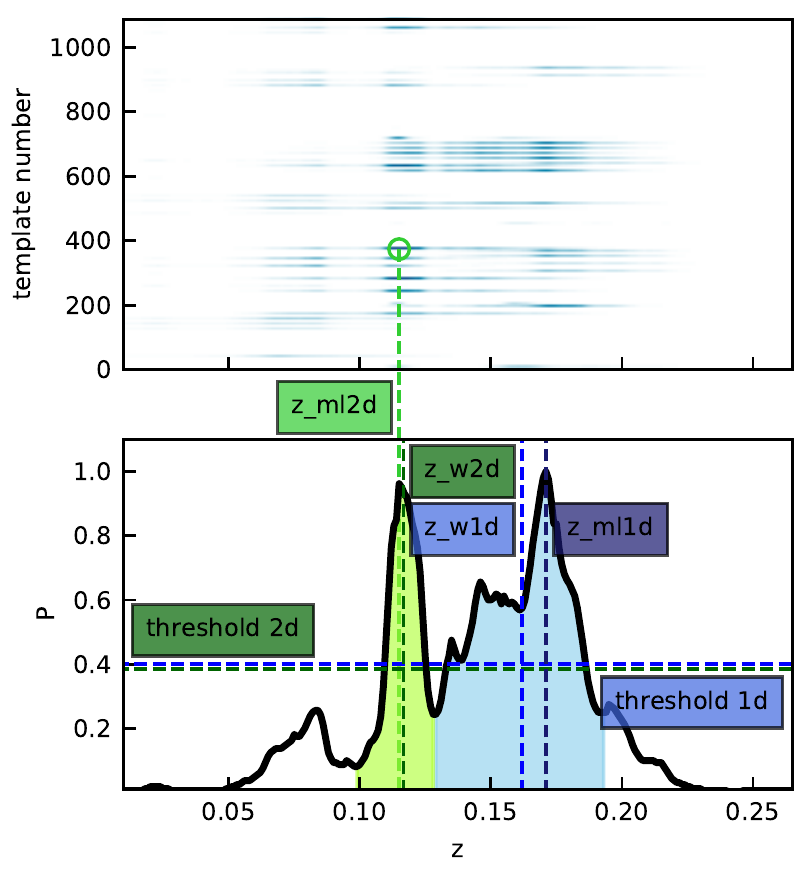}
    \caption{Example output of TOPz for one galaxy. The upper panel shows the $\chi^2$ heatmap of every template, where white and blue colours denote low and high likelihood, respectively. The green circle marks the location of the highest likelihood value. The lower panel shows the marginalised probability distribution. Different redshift estimators (described in Sect.~\ref{sec:benefits}) are marked with dashed vertical lines.}
    \label{fig:pdf_output}
\end{figure}

The general idea of Bayesian photo-$z$ codes is to calculate the redshift likelihood of a galaxy and estimate the `best guess' redshift from the respective probability density function (PDF).
This is illustrated in Fig.~\ref{fig:pdf_output} using an example of an actual galaxy from the miniJPAS survey. 
Typically, the redshift value corresponding to the highest PDF value is given as the `best guess' photometric redshift of the galaxy (designated {\tt z\_ml1d} in TOPz). 
Another possibility is to find the redshift with the highest likelihood value among all the given templates ({\tt z\_ml2d} in TOPz).

Depending on the observational uncertainties and the spectral characteristics of the galaxy, the PDF shapes can vary dramatically. 
Therefore, depending on the goal and on the actual data, other kinds of `best' redshift estimators can be used, for example taking into account the area under the likelihood curve up to some distance from the peak. Currently, two of such additional estimators have been implemented in TOPz, described below.

{\tt z\_w1d} is the weighted average of the PDF around the initial likelihood maximum. 
It is obtained by recognising the highest PDF value and then tracing the PDF peak in both directions until the first minimum below a user-defined threshold. 
The weighted average is then calculated over the traced part of the PDF. 
This redshift estimate performs best in situations where the peak of the PDF is not at the centre of a broader elevation as, instead of the peak location, the whole immediate area around the peak is taken into account.
Similarly, {\tt z\_w2d} is calculated using the redshift of the lowest $\chi^2$ value as the starting point and finding the weighted average over the traced part of the PDF.
In this paper, we have mostly used the {\tt z\_w1d} estimation as it was the best performing estimator on our test catalogue (see Sect.~\ref{sec:PDF_result}). A more thorough assessment of the performance of different `best' redshift estimators with different input data and in different redshift regimes has to wait for a larger data set from the upcoming J-PAS full survey.

For each photometric redshift value, we also give an `odds' estimate which is the relative area of the PDF within a user-defined fixed range centred on the estimated redshift value.
Odds value close to one means that the PDF is narrowly condensed around the highest PDF value whereas a low odds value means that the PDF is broad and the estimated redshift is of a lower probability.

In addition to the specific `best' redshift estimates, the full redshift PDF can be extracted as the output. 
In statistical analyses, the full posterior PDF (Eq.~\ref{eq:p}) gives a more adequate estimate of the spatial distribution of galaxies, for example for studies of clustering or the galaxy luminosity function \citep{2015MNRAS.452..549A, 2016MNRAS.456.4291A, 2017A&A...599A..62L}.
In the example shown in Fig.~\ref{fig:pdf_output}, the one-dimensional PDF has two separate peaks of roughly the same height. 
The redshift at the lowest $\chi^2$ value corresponds to one peak ({\tt z\_ml2d}) and the redshift at the highest value on the one-dimensional PDF corresponds to the other ({\tt z\_ml1d}).
The corresponding weighted averages are given with the dashed vertical, slightly darker, lines ({\tt z\_w1d} and {\tt z\_w2d}). 
The dashed horizontal lines represent the user-defined threshold for tracing the PDF peak.
In this figure, the threshold is set to 40\% of the peak value and the two thresholds are labelled {\tt threshold 1d} and {\tt threshold 2d} to note the two separate peaks of {\tt z\_ml1d} and {\tt z\_ml2d}, respectively.
The coloured areas indicate the traced parts of the PDF that are used to calculate the weighted averages.
In this specific case, the two peaks and the redshift estimations are all different, whereas in many cases they coincide.

\section{The miniJPAS catalogue}
\label{sec:mini_jpas}

MiniJPAS \citep{2021A&A...653A..31B} is a precursory photometric survey to the Javalambre Physics of the Accelerating Universe Astrophysical Survey \citep[J-PAS,][]{2014arXiv1403.5237B}.
It was carried out between May and September of 2018, using the 2.5-m Javalambre Survey Telescope, JST/T250, located in Sierra de Javalambre in Teruel, Spain. 
The observations consist of four pointings in the All-wavelength Extended Groth strip International Survey \citep[AEGIS,][]{2007ApJ...660L...1D} field covering a stripe of $\sim$1 deg$^2$ (1.9 deg $\times$ 0.5 deg).
The photometric data set was observed in 54 narrow-band and six medium and wide-band filters using the interim JPAS-Pathfinder camera. 
For our analysis, we have used the point-spread-function-corrected fluxes given in the {\tt minijpas.FNuDualObj} catalogue of the miniJPAS data release PDR201912, provided in the J-PAS data archive\footnote{http://archive.cefca.es/catalogues/minijpas-pdr201912}. 
Additional Milky Way extinction corrections have been applied per passband based on the E($B-V$) colour excess (see \citealt{2019A&A...631A.119L} for details). 

For characterising the brightness of the sources and applying luminosity cuts we have used the $r$-band {\tt MAG\_AUTO} magnitudes.
The specific luminosity cut values are described in the text.

The bulk of the science cases of the J-PAS survey critically depend on the accuracy of galaxy redshift measurements.
One of the most important constraints on the redshift accuracy is related to probing the properties of dark energy by measuring the scale of baryon acoustic oscillations \citep{2018MNRAS.477.3892C}.
Following these requirements, the target redshift accuracy of the J-PAS survey has been set to d$z/(1 + z) =|z_{\rm{phot}}-z_{\rm{spec}}|/(1 + z_{\rm{spec}}) < 0.003$ \citep{2014arXiv1403.5237B}. 
First proof that this target is reachable came with the miniJPAS data release, which includes galaxy redshifts derived with the LePhare code\footnote{http://www.cfht.hawaii.edu/$\sim$arnouts/lephare.html} \citep{2006A&A...457..841I}, specially modified to work with a larger number of filters and a higher resolution in redshift \citep{2021A&A...654A.101H}.
For example, for galaxies with $r<22$ mag, the normalised median absolute deviation of redshifts $\sigma_{NMAD}=0.0032\pm0.002$ was reached at a completeness level of 50\%  \citep[also see][for a more detailed assessment of the redshifts]{2021A&A...653A..31B}.

To test the accuracy of the photometric redshifts yielded by the TOPz workflow, we use a subset of miniJPAS objects that have reliable spectroscopic redshift estimates. 
The deep extragalactic evolutionary probe 2 (DEEP2) and deep extragalactic evolutionary probe 3 (DEEP3) Galaxy Redshift Surveys \citep{2011ApJS..193...14C,2013ApJS..208....5N} provide the largest and most comprehensive set of spectroscopic observations in the miniJPAS footprint. 
We only considered sources that are classified spectroscopically as a galaxy with the secure spectroscopic redshift quality flag (ZQUALITY $\ge 3$).
Some additional spectroscopic redshifts were cross-referenced from the Sloan Digital Sky Survey.
The depth of the test catalogue is set at $23$ magnitudes in $r$-band and the spectroscopic redshifts are selected up to redshift $1.5$.
In total, the resultant test catalogue consists of 4457 galaxies while a brighter ($r<22$ mag) sub-sample with 1989 objects is used in most of the tests.
In general, this test catalogue is based on the one used by \citet{2021A&A...654A.101H} for constructing the miniJPAS photo-$z$ and therefore we have considered the redshift accuracy achieved by them in the miniJPAS data release as a benchmark for TOPz performance.

Since each object in the test sample was required to have a spectroscopic classification of a galaxy, there are no objects in the test catalogue that are identified as AGNs or quasars, high-z point sources that exhibit distinct characteristics differing from other galaxies. Therefore, we have not attempted to represent AGN spectra in our templates.
We note that a robust detection and classification of J-PAS quasars would be needed to properly address them in the future \citep{2022arXiv220200103Q}.

\begin{figure}
    \centering
    \includegraphics[width=\linewidth]{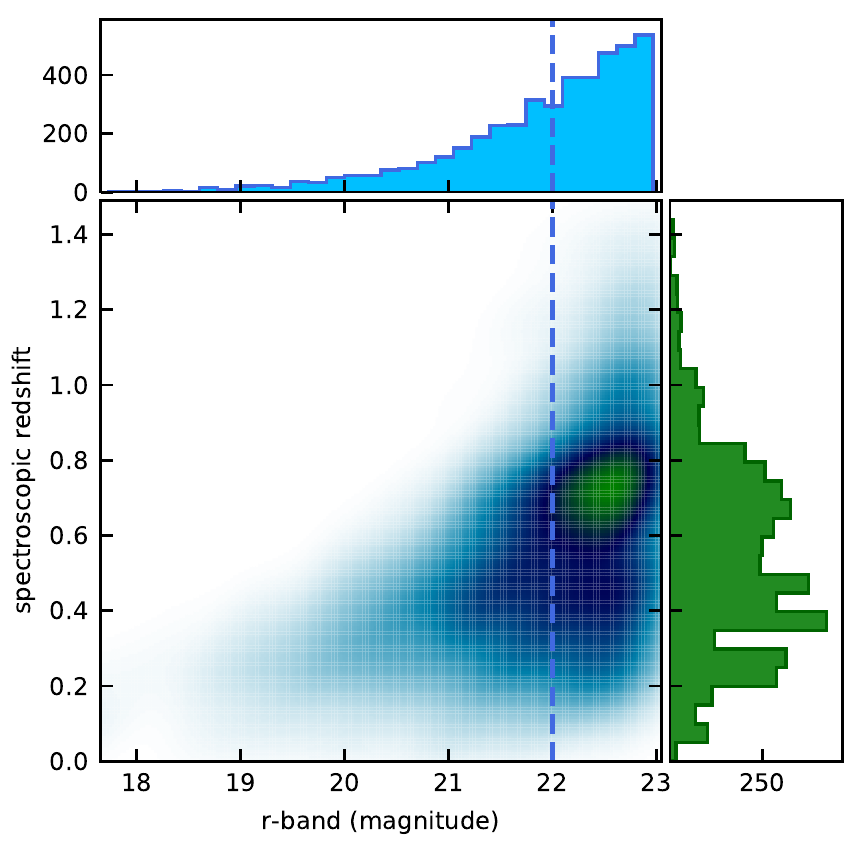}
    \caption{Distribution of spectroscopic redshifts and $r$-band magnitudes of the miniJPAS sources in the test catalogue. The dashed line represent the magnitude cut of the brighter sub-sample.}
    \label{fig:input_catalog_distribution}
\end{figure}

The distributions of spectroscopic redshifts and $r$-band magnitudes of the test catalogue sources can be seen in Fig.~\ref{fig:input_catalog_distribution}. 
The statistical fluctuations of redshifts are visible, hinting that a larger test sample would be desirable for a comprehensive assessment of the redshift estimates.
For example, the redshift histogram (green barplot on the right panel) shows peaks at around redshift 0.25 and 0.4 that are most likely due to larger clusters or superclusters in the line of sight of the observations.
Such concentrations may bias the final template selection.

The miniJPAS data reduction process identified some residual issues with individual images that needed special treatment: fringing, vignetting, and background patterns. 
To correct for these effects, additional illumination and background corrections were applied after the initial pipeline data reduction \citep[for more details see Appendix B of][]{2021A&A...653A..31B}.
Although these corrections generally improved the photometry, some systematic errors may have been introduced. 
Therefore, we seized the opportunity offered by the template-fitting redshift estimation principle to further refine the miniJPAS photometry. 
This step is described in detail in Sect.~\ref{sec:photo_corr}.

\section{Templates and prior of model galaxies}\label{sec:templates}

For template-based photometric redshift estimation, the quality of the template library is crucial.
Moreover, depending on the observational data and the aims, specific templates with specific features may be needed.
For example, the spectral range and spectral resolution of the observational data set and the targeted redshift range dictate the requirements for the spectral range and spectral resolution of the templates.
Additionally, the presence and precision of broad absorption features such as MgB, CaT, or TiO bands and emission lines in the templates are needed. 
Besides the technical characteristics of the templates themselves, the whole template library must be representative of the observed sources. 
Counter-intuitively, a maximally broad choice of templates is often not the best solution because at lower signal-to-noise levels the chance of a completely unrealistic template yielding a low $\chi^2$ value at an incorrect redshift increases significantly.

The J-PAS optical system comprises of observations that range from $3400$ to $11\,000$\:\AA.
Since several sources are expected to lie well beyond $z=1$, we require the templates to cover at least the $1000$ to $11\,000$\:\AA\:range. 
The minimally required spectral resolution is not determined by the width of the passbands (FWHM $\sim145$\:\AA) but rather by the sharp edges of the throughput curves.
These edges are measured to be around $10$\:\AA\:wide, so a template spectral sampling rate of $\leq10$\:\AA\:would be needed.

Generally, there are two approaches to compile the template library: using observed spectra of a broad range of galaxy types or using stellar population synthesis models. 
Since there are only a small number of readily available observational spectra covering the required wavelength range at the required resolution, we consider the synthesis approach to be more applicable. 
Besides, the spectra produced by stellar population synthesis can be directly linked to the physical properties of the galaxies, making it easy to provide a value-added catalogue alongside with the redshift estimates.

As mentioned above, we would like to prevent the template library from becoming too large. 
On the other hand, the usage of narrow-band photometry requires the templates to be realistic and representative also in details like spectral line strengths and ratios. 
We chose the strategy of constructing templates on the basis of the same data set which the templates are going to be applied on, in the present case, the miniJPAS \citep{2021A&A...654A.101H}. 
While risking to introduce a biased choice of templates due to the small number of objects in the data set, we are at least ensuring that the template library corresponds to the actually targeted sources.

\subsection{Base template set generation}
\label{sec:template_generation}

In this work, we used the Code Investigating GALaxy Emission \citep[\textit{CIGALE};][]{2019A&A...622A.103B} for generating a library of synthetic spectra. 
\textit{CIGALE} uses filter passbands and observed galaxy fluxes through these passbands to generate its synthetic spectrum.
Assuming the spectroscopic redshift of an observed galaxy, we could construct precise synthetic galaxies from the spectroscopically observed subset of the full miniJPAS data set (see Sect.~\ref{sec:mini_jpas}). 
We varied the \textit{CIGALE} input parameters determining the star formation history, dust attenuation, and other properties within realistic limits. 
According to its working principle, \textit{CIGALE} calculates the spectra resulting from each parameter combination and then assesses their correspondence to the observed galaxy's SED.
Eventually, we are left with \textit{CIGALE}'s best estimate for the spectrum of each galaxy. 
In order to avoid approximating noisy data with unrealistic models, we used 500 brightest galaxies from the test sample for this procedure, that is galaxies up to $\sim21$ mag in $r$-band. 
We fine-tuned almost every \textit{CIGALE} input parameter to achieve the best TOPz photo-$z$ accuracy of the sub-sample. 
The list of input parameters generating the template library used throughout this work is specified in Table~\ref{tab:cigale_param}.

\begin{table}
    \centering
    \caption{Parameter names in \textit{CIGALE} alongside the value ranges that were used to construct the templates.}

    \label{tab:cigale_param}
    \begin{tabular}{ll}
        \hline\hline
        Parameter & Values \\
        \hline

        \multicolumn{2}{c}{Delayed exp\tablefootmark{a} with burst (\textit{sfhdelayedbq} module)}\\
        \rule{-2pt}{2ex}

        $\tau_{\rm main}$\tablefootmark{b} & 300, 680, 1550, 3500, 8000 \\
        Age of main population &  2000, 5300, 8600, 12000\\
        Age of burst & 10, 29, 83, 240, 690,  2000 \\
        Ratio of the SFR\tablefootmark{c} after burst & 0.01      ,  0.017,  0.03,  0.05,  0.08, \\ &
        0.14,  0.24,  0.4  \\

        \multicolumn{2}{c}{Bruzual Charlot synthesis (\textit{bc03} module)}\\
        \rule{-2pt}{2ex}
        IMF\tablefootmark{d} & Salpeter \\
        Metallicity & 0.008, 0.02\\
        Population age separation & 10, 32, 100, 316, 1000\\
        %
        \multicolumn{2}{c}{Emission features (\textit{nebular} module)}\\
        \rule{-2pt}{2ex}
        Ionisation parameter &  $-3.0$, $-2.5$, $-2.0$, $-1.5$, $-1.0$ \\
        %
        \multicolumn{2}{c}{Dust absorption (\textit{dustatt$\_$modified$\_$CF00} module)}\\
        \rule{0pt}{2ex}
        $\mu$\tablefootmark{e} & 0.01,  0.025,  0.06,  0.16,  0.4,  1\\
        ISM\tablefootmark{f} power law slope & $-0.5$, $-0.7$, $-0.9$\\
        ISM $V$-band attenuation  &  1\\
        Power law slope of the  & \\
        attenuation in birth clouds &$-1.3$\\
        
        \hline

    \end{tabular}
    \tablefoot{Abbreviations:
    \tablefoottext{a}{exponential model;}
    \tablefoottext{b}{e-folding time of the main stellar population model in Myr;}
    \tablefoottext{c}{star formation rate;}
    \tablefoottext{d}{initial mass function;}
    \tablefoottext{e}{attenuation ratio;}
    \tablefoottext{f}{interstellar medium.}
    }

\end{table}

From Fig.~\ref{fig:input_catalog_distribution}, it can be seen that the 500 brightest galaxies that we used for the template construction have redshifts $z\lesssim0.7$ while the brighter test sample extends farther, $z\lesssim1$.
However, since  evolutionary effects become important at much higher redshifts, it is very unlikely that there are galaxy types in our test catalogue that are not represented by the 500 brightest galaxies and therefore no additional templates are needed to compensate for the redshift difference.
A test on how under or over-represented templates affect the resulting redshift estimations is discussed in Sect.~\ref{sec:test_templates}.

The resulting spectra were designated as our base set of templates.
In principle, all of these spectra can be used as templates, but initial tests suggested that this is not optimal for photometric redshift determination. 
In order to mitigate this issue, we selected a template sub-sample that would yield a better photometric redshift accuracy as described in the next Section.

\subsection{Final template selection}
\label{sec:template_selection}

In addition to the template quality and representativeness, also the size of the template library can be optimised. 
A higher number of templates increases the redshift accuracy of brighter sources, while also lowering the accuracy of fainter ones due to over-fitting. 
Therefore, the optimal number of templates for different brightness sub-samples are different.
The results of the corresponding test are described in Sect.~\ref{sec:test_templates}. 
For the brighter ($r<22$ mag) subset, we find that a library of around 75 templates gives an optimal result.

In order to obtain a smaller template library while still providing an adequate description of the observed galaxies, we reduced our base template set on the basis of minimising the total $\chi^2$ value.
In essence, the selection process was done by selecting a specified number of templates from the base template set, shifting them by the spectroscopic redshift value, and then applying the TOPz workflow to the test catalogue.
While iteratively selecting different sets of templates, we would find the one that minimises the total $\chi^2$ value of the test catalogue.

A more detailed description of the template selection process is as follows.
Similarly to Eq~\ref{eq:chi2}, we calculated the $\chi_{\rm T}^2$ for each galaxy-template combination as
\begin{equation}
\chi_{\rm T}^2 = \frac1{N_{\rm filt}}\sum\limits_{j = 1}^{N_{\rm filt}} \left(\frac{F_j - aF_{T,j}}{\sigma^2_{F_j}}\right)^2,
\end{equation}
where $N_{\rm filt}$ is the number of valid pasbands for a given galaxy.

Let us denote the set of selected templates as $V$ so that each set $V$ belongs to the overall set containing all possible combinations $V\subset V_{\rm master}$ while the number of templates in set $V$ is fixed at $N_{\rm T}$. 
For every galaxy, we selected the best-matching template and the corresponding $\chi^2_{\rm T}$ value from $V$ by calculating
\begin{equation}
    \chi^2_{\rm gal} = \min_{T \in V} \left( \chi^2_{\rm T} \right).
\end{equation}
From there, the general assessment of the template set $V$ was based on the total $\chi^2_{\rm cat}$ value of the test catalogue (with $N_{\rm gal}$ galaxies) by calculating
\begin{equation}
    \chi^2_{\rm cat} = \sum\limits_{i = 1}^{N_{\rm gal}} \chi^2_{\rm gal} w_{\rm gal},
\end{equation}
where the weights $w_{\rm gal}$ are based on the normalised apparent magnitude of the galaxy. 
To lower the effect of the more noisy fainter galaxies, the weight values used in this paper were between $3$ and $1$.
This means that the brightest galaxy has a weight value of $3$ and the faintest galaxy a weight value of $1$ with other galaxies having weight values linearly distributed between those two.
This ensures that when the $\chi^2_{\rm gal}$ values change between different template sets $V$, the brighter galaxies would contribute more to the total $\chi^2_{\rm cat}$ change than the fainter galaxies.

The selection of the best template set was calculated iteratively. 
The first iteration contained a random initial set of templates while every other iteration replaced a single template, starting from the worst performing one.
For determining the best template set, $V_{i}$ was substituted with $V_{i+1}$ only in case the global $\chi^2_{\rm cat}$ is minimised ($\chi^2_{{\rm cat}, i+1}< \chi^2_{{\rm cat},i}$).
Otherwise, the set $V_{i}$ would be the starting point for the next iteration.

\begin{figure}
    \centering
    \includegraphics[width=\linewidth]{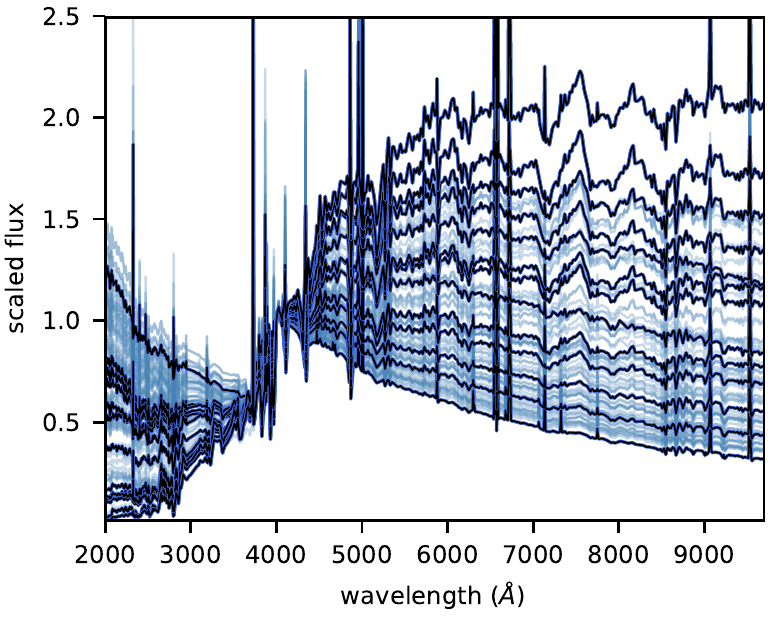}
    \caption{Final selection of 75 templates after pruning the base template set made with \textit{CIGALE}. The fluxes are scaled to unity at 4000\:\AA. Some templates are depicted darker for visual representation.}
    \label{fig:all_templates}
\end{figure}

\begin{figure}
    \centering
    \includegraphics[width=\linewidth]{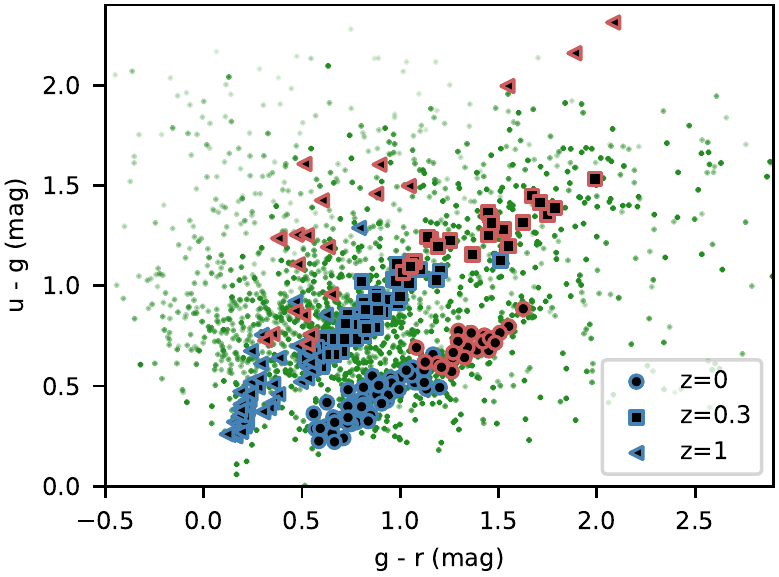}
    \caption{Colour-colour diagram of the miniJPAS catalogue (green points) and the selected templates (different markings). Green point intensity corresponds to the galaxy $g$-band flux. The shapes of the template markings indicate the same templates at three distinct redshifts. Blue and red colours indicate whether a late- or early type prior was applied to the given template (see Sect.~\ref{sec:prior}).}
    \label{fig:templates_colours}
\end{figure}

Following this method, we reduced the total number of templates while covering most of the spectral features present in the SEDs of the galaxies in our test catalogue.
All of the 75 templates in the final selection are plotted in Fig.~\ref{fig:all_templates}.
The fluxes are scaled at 4000\:\AA\:in order to see the shapes of the templates. 
In this selection, there are templates with and without emission lines although it is hard to see on the plot as the emission lines overlap with the templates without the lines.

Another way of showing the template coverage is using a colour-colour diagram (see Fig.~\ref{fig:templates_colours}).
The $r<22$ miniJPAS catalogue until $z=1$ are shown as green points while fainter points indicate fainter galaxies in $g$-band flux.
The template colours are shown for three distinct redshift values: 0, 0.3, and 1. 
Increasing redshift values twist and shift the template colours towards the upper-left side of the colour-colour diagram.
As such, the area between the $z=0$ and $z=1$ template colours is where we expect most of our redshifted templates to lie.
The overlapping region between the template colours and catalogue colours shows where our selected templates cover the corresponding galaxy types.
We define a rectangular region between the templates with most extreme colours as the colour range that is encapsulated by the templates.
Most of the catalogue galaxies ($\sim 78\%$) are within this region, while there are some galaxies that have more extreme $g-r$ colours than our templates.
These measured colour extremes are likely due to fainter galaxies as most of the faint (less $g$-band flux) green points lie in that region.

\subsection{Photometric corrections}
\label{sec:photo_corr}

By knowing the spectroscopic redshifts and the expected (template) spectra of the galaxies, we can inspect the observed photometry for possible systematic offsets. 
To check for such offsets, we used the template selection as described in Sect.~\ref{sec:template_selection}.

\begin{figure}
    \centering
    \includegraphics[width=\linewidth]{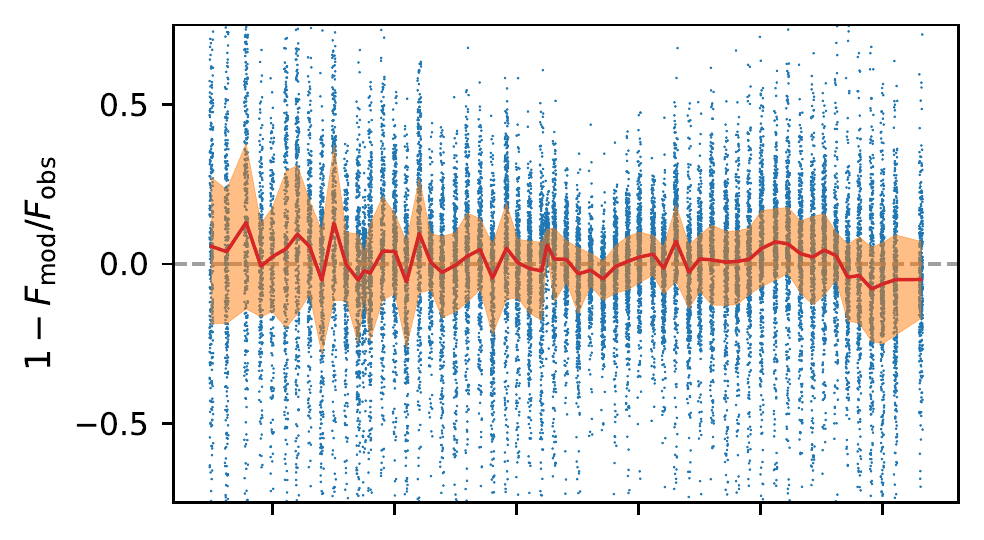}
    \includegraphics[width=\linewidth]{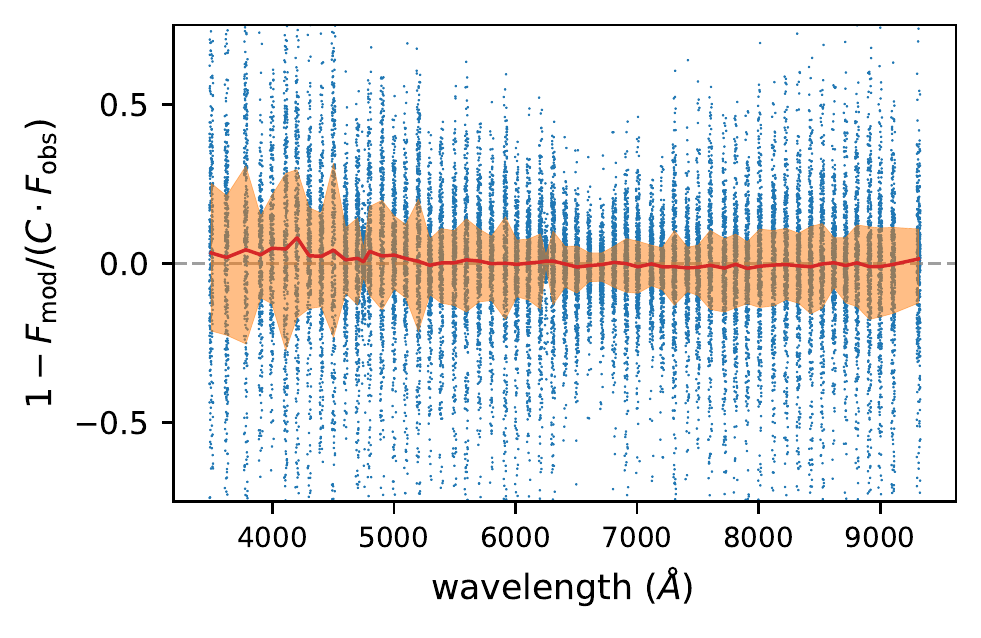}
    \caption{Offsets between the observed and template-based model fluxes before (upper panel) and after (lower panel) the photometric corrections. The correction factor $C$ is defined in Eq. \ref{eq:phot_corr}. Solid line illustrates the median value while the shaded area marks the inner 50\% of the objects in each passband.}
    \label{fig:obsmod_diff}
\end{figure}

For every galaxy, we calculated the difference between the observed and synthetic photometry that is obtained from the template with minimum $\chi^2$ value at the spectroscopically fixed redshift.
The upper panel of Fig.~\ref{fig:obsmod_diff} presents such differences for each passband. 
While the scatter is large, a notable systematic offset of median values (up to $10\%$) is also present in many filters. 
Assuming that the templates at least roughly depict the actual galaxy SEDs, we may consider that systematic offsets between observations and templates are unlikely to be caused by problems with the templates.
This is because the offsets occur even after redshifting the templates by a varying amount according to each given source and therefore should not be a systematic effect caused by the templates.
Instead, the offsets refer to some residual deviations in the photometry that we can reduce by bringing the observed photometry closer to the templates, as done in \citet{2006AJ....132..926C}.
We also note that this kind of correction might introduce some colour terms, that is correlations between passbands, as synthetic spectra do not represent all the aspects of observed galaxies. 
So one has to keep in mind that these corrections may still contain some dependence on the template or source set and might not be applicable for other purposes.

For calculating the photometric corrections, we also considered the known observational uncertainties.
The correction term is defined so that the average difference between the observations and synthetic photometry would become zero.
For each passband, the correction term C is calculated using the following expression:
\begin{equation}
    \sum_{i=1}^{N_\mathrm{gal}} \frac{F_{T,i} - C F_i}{\sigma_{F_i}} = 0,\
\label{eq:phot_corr}
\end{equation}
where $F_{T,i}$ and $F_i$ are the synthetic and observed fluxes of each galaxy $i$, and $\sigma_{F_i}$ are the corresponding observational uncertainties.
Factor $C$ is the correction term for the given passband that is set to one for uncorrected data and differs from unity if correction is needed.
After the initial run, we applied the corrections to the observations in each passband and conducted another iteration of TOPz with the newly corrected photometry while keeping the same templates. 
We iterated up to four times until no significant improvement could be seen between the last two iterations; final correction value would thus be the cumulative correction over the iterations.
While correcting the observations, we kept the observational error at the same fractional value that it was in the original catalogue.
This means that when the brightness increased due to photometric corrections, the absolute observational errors were also increased and vice versa.

The miniJPAS observations consist of four tiles, that is co-added frames of multiple exposures observed through each of the J-PAS filters at four different pointings. 
We calculated the correction terms on a per tile basis using only the galaxies in said tile.
We found slightly different correction terms for each tile that are most likely associated with the variable PSF of the images that comprise these tiles.
Although the correction terms from tile to tile might differ a lot in some of the passbands, the overall shape of the terms in all of the passbands remain the same.

\begin{figure}
    \centering
    \includegraphics[width=\linewidth]{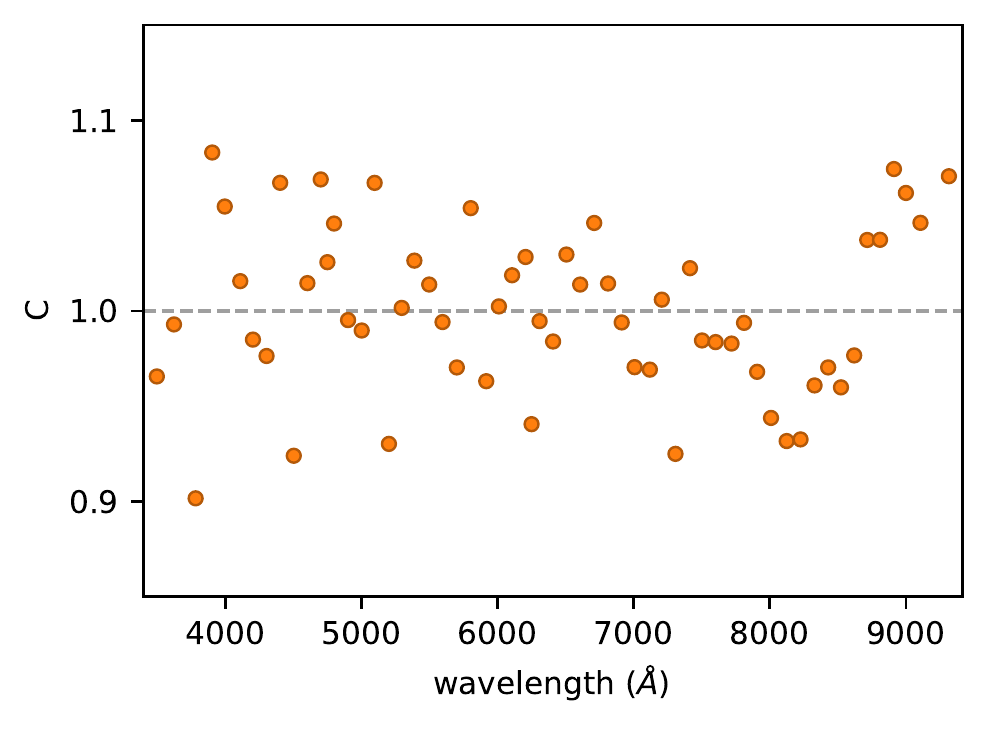}
    \caption{Photometric correction term $C$ (defined in Eq.~\ref{eq:phot_corr}) for each passband given in wavelengths. This is the average correction value over the four different tiles in miniJPAS data.}
    \label{fig:zp_corrections}
\end{figure}

Figure \ref{fig:zp_corrections} shows our calculated corrections in all of the miniJPAS passbands where each point denotes the correction term that is averaged over the four different tiles for clarity.
As can be seen, the correction values range from 0.9 to 1.1 and are mostly uncorrelated with passband wavelength.
An exception is the red end (starting from $\sim8000\:\AA$) where a sharp gradient is seen. 
We speculate that this could be related to the fringing effect affecting the CCD photometry at longer wavelengths.
It also seems that the average correction term between $4000$ and $7000\:\:\AA$ trends towards unity, which is probably related to the decrease in observational errors caused by the atmospheric absorption in the `optical window'. 
The corrected photometry can be seen on the lower panel of Fig.~\ref{fig:obsmod_diff} which shows the differences between observational and model fluxes after applying the photometeric corrections from Fig.~\ref{fig:zp_corrections} to the observations. 
In general, the average difference between observations and model has been reduced significantly, while some minor systematic offsets are still present at bluer wavelengths.

\subsection{Prior}
\label{sec:prior}

The prior, in the context of Bayesian redshift estimation, enables us to refine our results using what we know about the distribution of galaxy luminosities and galaxy types at different redshifts.
The prior indicates the probability of finding a galaxy with a certain apparent magnitude and type at a certain redshift. 
This probability can be inserted into the redshift estimation through the prior term $p(z\,|\,T,m_0)$ in Eq.~\eqref{eq:prior} modifying the resultant redshift PDF.

\begin{figure}
    \centering
    \includegraphics[width=\linewidth]{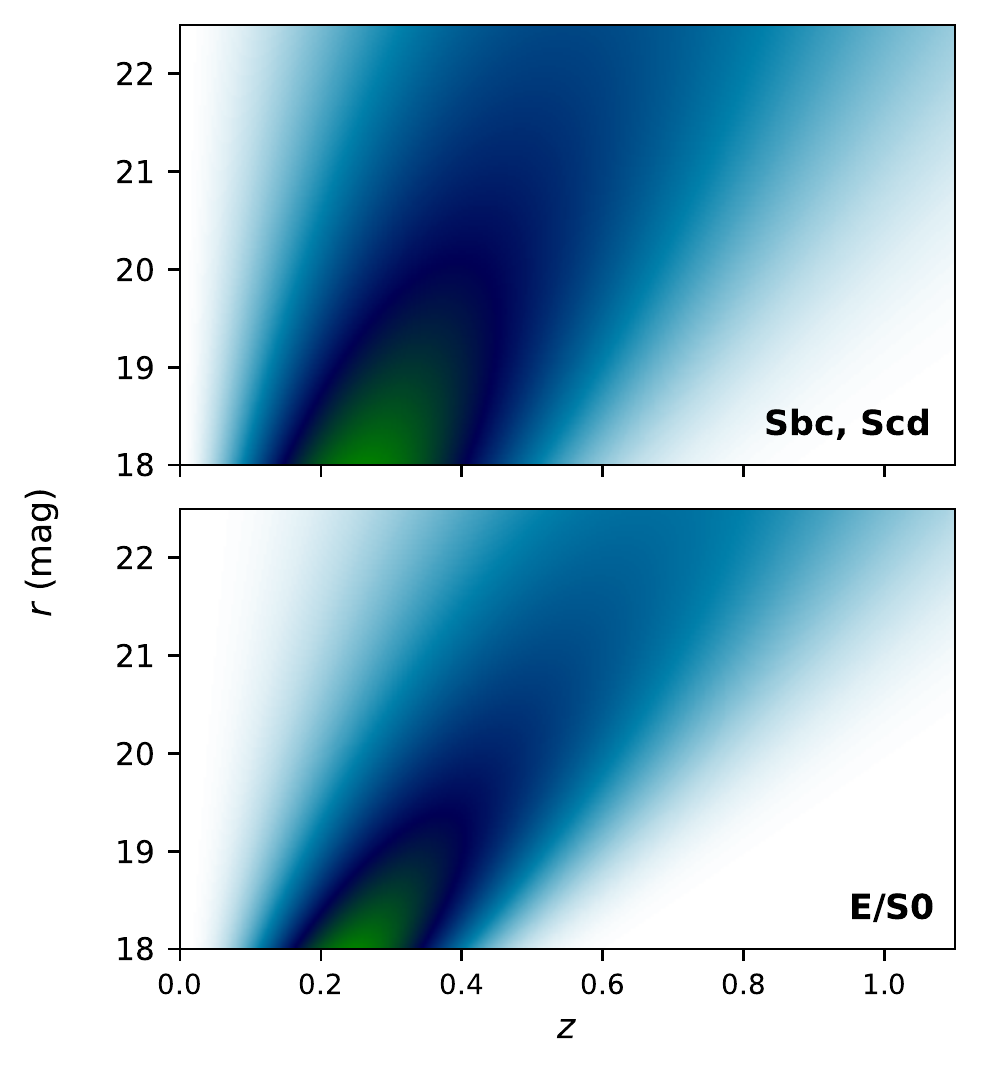}
    \caption{Two types of galaxy priors used in this work. The upper panel shows the probability that a spiral galaxy of a certain brightness would lie at a certain redshift and the lower panel shows the same for an early type galaxy.}
    \label{fig:prior_comp}
\end{figure}

In principle, it is possible to ascribe a different prior to each template and, provided that the galaxy data set under consideration is large enough, the priors for each template can be calibrated based on the  data themselves.
However, given that our current sample is rather limited and external observations do not give information at the required level of detail, we have adapted here the priors constructed by \citet{2000ApJ...536..571B}. 
He introduced an analytic form for the prior, containing a parametrised relationship between the redshift and apparent luminosity, depending on the morphological type. 
Three morphological types were distinguished: early type galaxies (E/S0), spirals (Sbc, Scd), and irregulars. 
The parametrisation of the prior function was conducted on the basis of galaxy statistics in the Hubble Deep Field North \citep{1996AJ....112.1335W}.

We split our templates into red and blue sub-types according to their cumulative spectral distribution and respectively ascribed them the prior function of either the early type or the spiral type from \citet{2000ApJ...536..571B}. 
That is, we calculate the wavelength value where the cumulative template flux reaches 50\% and then apply the spiral type to $\sim80\%$ of galaxies with lower wavelength value and early type to the rest, roughly following the  fractions measured for the local universe.
The colour-colour representation of the sub-types can be seen in Fig.~\ref{fig:templates_colours} where, depending on the redshift, each group of templates is separated into a redder and bluer marking denoting the red and blue sub-types, respectively.
The dependence on the redshift for the early and spiral type priors differ mostly when looking at fainter galaxies (see Fig.~\ref{fig:prior_comp}).
We chose the $r$-band magnitude as an indicator of the apparent luminosity. 
Our tests showed that the eventual redshift accuracy was not sensitive to the exact location of separation between red and blue galaxies.

\begin{figure}
    \centering
    \includegraphics[width=\linewidth]{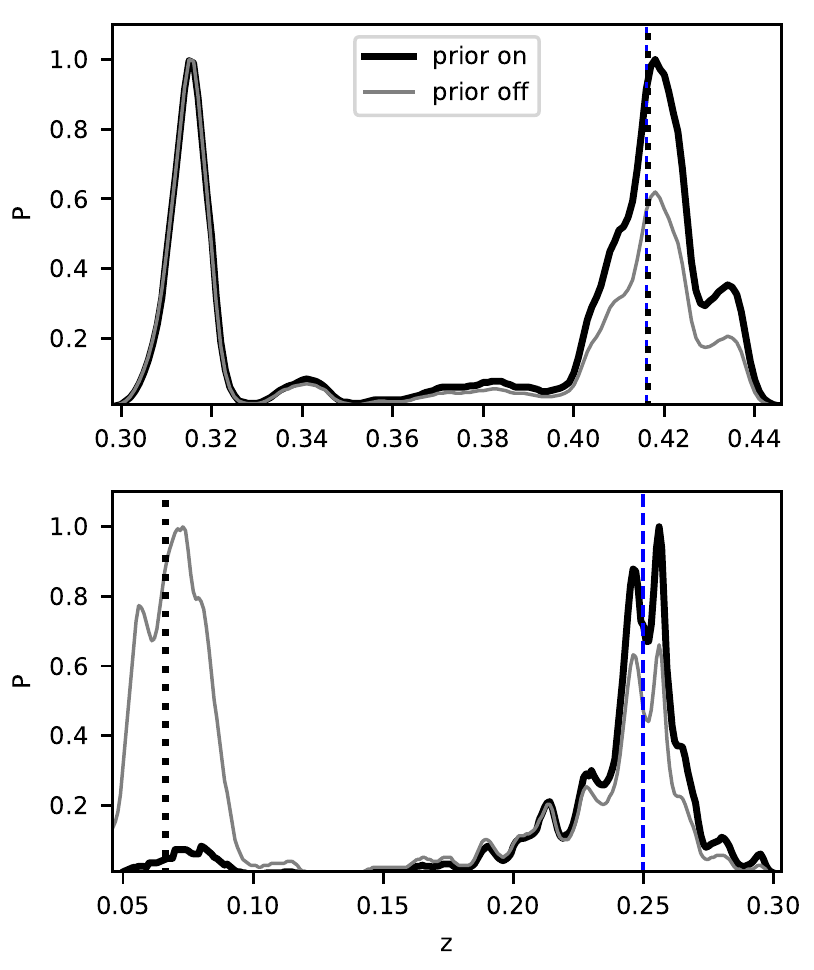}
    \caption{Effect of prior on the redshift PDFs of two selected galaxies. The black dotted vertical line is the spectroscopic redshift and the dashed blue line is the TOPz {\tt z\_w1d} redshift estimation. Solid black and grey lines show the normalised PDFs when considering or not considering the prior, respectively.}
    \label{fig:posteriors_prior_comp}
\end{figure}

The prior bends the redshift PDF according to the overall shape of the prior (see Fig.~\ref{fig:posteriors_prior_comp}).
These drastic examples illustrate how adding a prior can both improve as well as hamper the redshift determination when the PDF has multiple peaks.
The prior can enhance the secondary maximum into a primary one and the resultant redshift value estimation become very close to the spectroscopic redshift value (see top panel of Fig.~\ref{fig:posteriors_prior_comp}).
Whereas, on the lower panel the prior drastically reduces the primary maximum that was centred around the spectroscopic redshift, giving the galaxy a wrong photometric redshift estimation.
This effect is most prominent for faint galaxies at very low redshift values where the prior favours a higher redshift solution.
Thus, there is room for improvements in specifying the prior by adding additional prior types and, if sufficient J-PAS data with known spectroscopic redshifts will be available, by constructing the priors directly from observations.

\section{Photometric redshifts with miniJPAS data}
\label{sec:minijpas_z}

For the tests with the miniJPAS data, the upper limit of TOPz redshift estimations were set to $z = 1.5$ and the resolution was set to $\Delta z = 0.001$.
The threshold value for calculating weighted estimates (see Fig.~\ref{fig:pdf_output}) was set to $0.4$ as the closest rounded value that, based on our initial testing, yielded the best results.
In addition, we applied the prior and the photometric corrections as described in Sect.~\ref{sec:photo_corr} and \ref{sec:prior}. 
Below we give an overview of the impact that different inputs and configurations have on the TOPz performance for the $r<22$ mag sub-sample of galaxies (1989 sources) when comparing the photometric redshift estimations to the spectroscopic redshifts.

\subsection{Testing the impact of TOPz inputs}
\label{sec:prior_template}

\subsubsection{Impact of template selection}
\label{sec:test_templates}

\begin{figure}
    \centering
    \includegraphics[width=\linewidth]{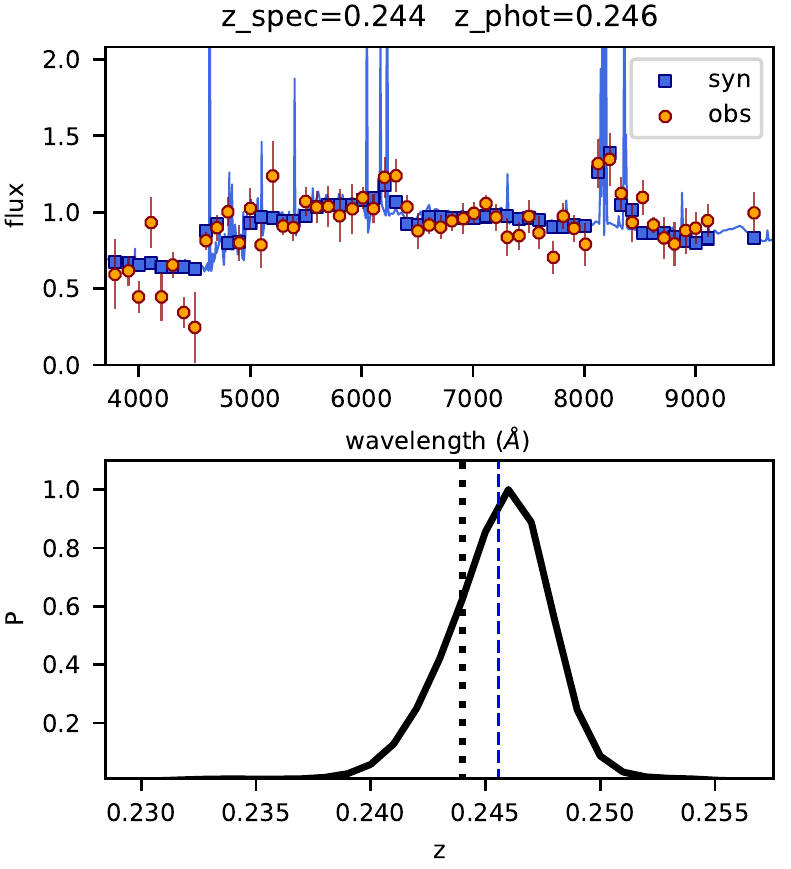}
    \caption{Example of a fitted template matching the photometry. The upper panel shows the match between the observed (orange circle) and the synthetic photometry (blue box) of the best template (blue line). The lower panel is the PDF from TOPz with dashed blue line showing the weighted redshift ({\tt z\_w1d}) and black dotted line the spectroscopic redshift.}
    \label{fig:spec_photo}
\end{figure}

As explained in Sect.~\ref{sec:templates}, photometric redshift estimation may depend on the set of templates used for approximating the observed spectral distribution of the galaxies. 
Consider the example of the best-matching template fitted to the photometry of an $r=20.6$ galaxy, presented in Fig.~\ref{fig:spec_photo}. 
On the upper panel, the blue line and squares represent the template spectrum and the corresponding synthetic photometry, respectively, and the orange circles are the observed fluxes together with error estimates in each of the 54 narrow-band filters.
Although the photometric errors are relatively large and the scatter of the observations even exceed these errors, we can quite accurately detect the major emission lines. 
The lower panel shows the marginalised PDF that is produced by our final template set. 
The PDF peak ({\tt z\_ml1d}) as well as the {\tt z\_w1d} redshift estimation (dashed blue line) are somewhat overestimated.
Nevertheless, the photometric redshift is more accurate than the J-PAS target goal of d$z/(1 + z) < 0.003$.

In Sect.~\ref{sec:template_selection} we noted that $\sim22\%$ of the galaxies in the test catalogue fall outside the colour region that our templates cover.
We also noted that these galaxies are fainter on average, having a median brightness of $r = 21.57$ mag compared to $r = 21.21$ mag of those galaxies that are inside the region.
We find that at a fixed brightness level, the number of galaxies that reach the J-PAS accuracy goal is similar between galaxies outside the colour region and the remaining galaxies.
This shows that, although the broadband colours of the templates are somewhat more restricted than those of the observed galaxies, the templates are accurate enough to yield reliable redshift estimates from the full J-PAS filter set.
The most probable explanation is that the accuracy of photo-z for fainter galaxies is, due to their larger photometric uncertainties, mostly defined by the detection of emission lines and not the template broadband colours themselves.

In TOPz, the marginalised PDF of a galaxy is determined by the input templates since it is combined from the $\chi^2$ values of every single template in that set.
If the chosen templates do not represent the observed data, the combined PDF shape will be worsened by the unsuitable templates, which may yield a relatively low $\chi^2$ fit to the observational data points at an otherwise random redshift.
Thus, by reducing the total number of templates, we can find a template configuration that improves the overall quality of the photometric redshifts compared to an arbitrarily composed big template set.
A more general solution could be made by applying a template prior to the likelihood.

First, we tested how varying the amount of a single template type affects the marginalisation process of the PDF and what is the effect on the resulting redshift estimations.
For this, we increased the ratio of templates that were defined as elliptical up to 50 times, increasing the template set from 75 up to $\sim1100$.
The resulting photometric redshift estimations worsened at most 1\% and thus we conclude that the number of templates of any certain spectral type has negligible effect of the outcome.
This is mainly because the ratio of templates with different spectral types only affect redshift estimations for galaxies with PDFs having multiple peaks (75\% of the test catalogue is single peaked) with sufficient strength. 
Therefore, a potential template bias that would be caused by the fact that DEEP2 and DEEP3 spectroscopic samples were selected to prefer higher-redshift sources should not affect our results.
This type of template selection effect could be further reduced by introducing a template prior that would weight each template type by its occurrence in the selection.

\begin{figure}
    \centering
    \includegraphics[width=\linewidth]{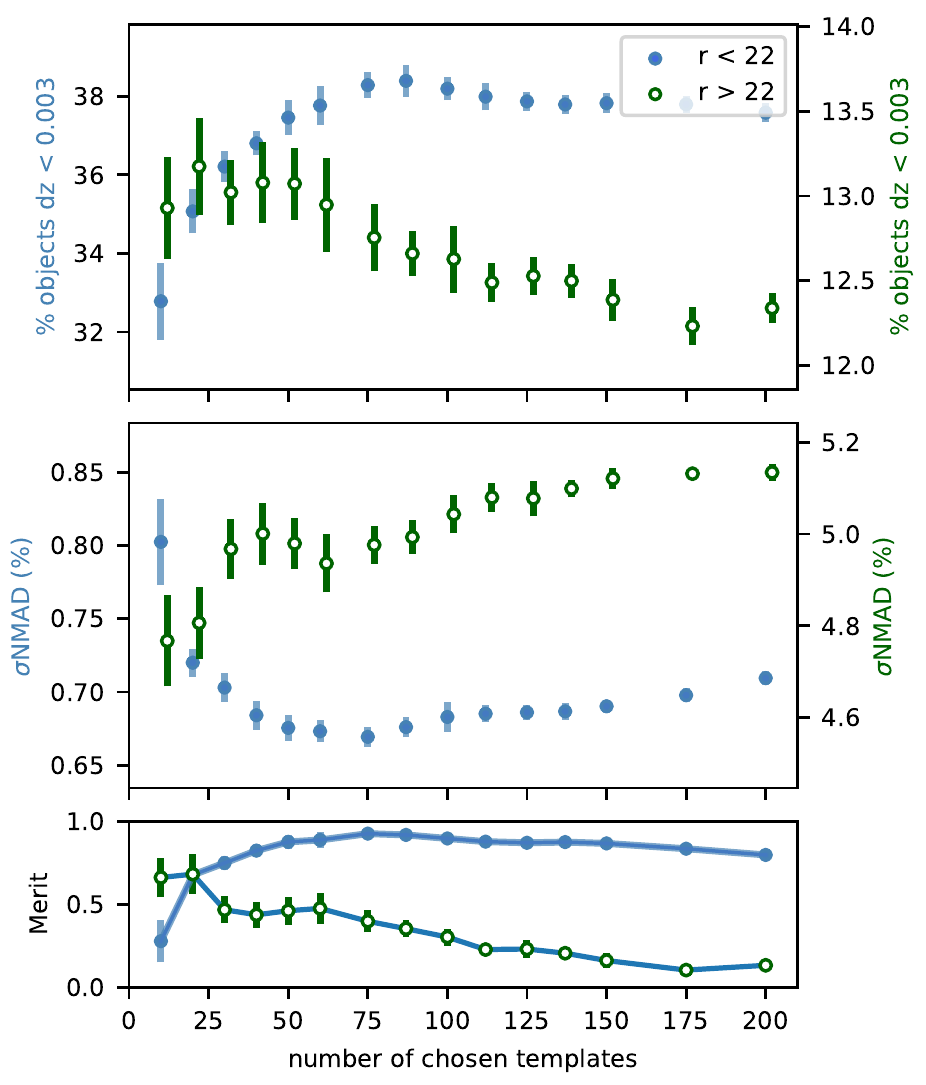}
    \caption{Photometric redshift accuracy depending on the size of the used template set. Each point marks the mean value of the 20 different realisations of a template selection run. Upper panel shows the percentage of objects that fit the J-PAS accuracy criteria while the middle panel shows the achieved median absolute deviation for the normalised distribution ($\sigma$NMAD). The lower panel shows a merit function for choosing the final number of templates. The two colours are cuts in magnitude where the green points are shifted slightly right for visibility.}
    \label{fig:template_grid}
\end{figure}

Next, we conducted a test to see the impact that the size of the template set has on the eventual redshift accuracy.
The results are shown in Fig.~\ref{fig:template_grid}. 
Each point represent 20 realisations of the template set selection run. 
The errors of these points are a result of the semi-random nature of the template selection procedure (see Sect.~\ref{sec:template_selection}).
The two colours show the results separately for objects brighter (blue) and fainter (green) than $r = 22$ mag.
The upper panel shows that the fraction of brighter galaxies that achieve the J-PAS target accuracy increases until the template set size of about 75 is reached.
This is because too few templates cannot cover the whole spectral type distribution of the observations.
Beyond the 75 template mark, additional templates do not improve the results. 
While more templates may provide a better approximation for some galaxies, they contaminate the redshift PDF of some others and effectively reduce the overall redshift determination accuracy. 
The middle panel of Fig.~\ref{fig:template_grid} shows a similar result for brighter galaxies when using the normalised median absolute deviation ($\sigma_{\rm NMAD} = 1.4826*median(|dz- median(dz)|)$ ) as a proxy to describe the spread of the photometric redshift accuracy \citep{2021A&A...654A.101H}. 
The deviation for the brighter sub-sample stays the lowest when the size of the template set is close to 75.

As Fig.~\ref{fig:template_grid} shows, the above aspects are different for fainter galaxies which are more dominated by noise.
Estimation of the redshifts of fainter galaxies benefits from the use of far fewer templates as then the possibility that an arbitrary template falsely gives a relatively low $\chi^2$ at an arbitrary redshift decreases.

Based on the above information as well as on the outlier rate, defined as galaxies with a redshift accuracy worse than 5\%, we have constructed a merit function that helps us choose the optimal number of templates.
We define the following counts:
\begin{eqnarray}
    U &=& \sum\limits_{i = 1}^{N_{\rm gal}}\mathbbm{1}\left[ \frac{dz_i}{1 + z_i} < 0.003\right],\\ 
    W &=& \sum\limits_{i = 1}^{N_{\rm gal}}\mathbbm{1}\left[ \frac{dz_i}{1 + z_i} < 0.05\right] ,
\end{eqnarray}
where U and W are the number of galaxies that reach the J-PAS accuracy limit and the number of galaxies that don't reach the outlier limit, respectively.
The merit function is then defined as a weighted combination of the $\Omega$ components:
\begin{equation}
    M_{\rm erit} = w_{\rm acc} \Omega_{\rm acc} + w_{\rm NMAD}  \Omega_{\rm NMAD} + w_{\rm out}  \Omega_{\rm out},
    \label{eq:merit}
\end{equation}
where each $\Omega$ component is defined as the corresponding values normalised between [0,1] 
\begin{eqnarray}
    \Omega_{{\rm acc},m} &=& \frac{U_{m} - \min{U}}{\max{U} - \min{U}}, \\
    \Omega_{{\rm NMAD},m} &=& \frac{\sigma_{{\rm NMAD},m} - \min{\sigma_{\rm NMAD}}}{\max{\sigma_{\rm NMAD}} - \min{\sigma_{\rm NMAD}}},\\
    \Omega_{{\rm out},m} &=& \frac{W_{m} - \min{W}}{\max{W} - \min{W}}.
\end{eqnarray}
Here, $U_m$ is the $U$ value for an applied template set $V_m$ and the same holds true for $\sigma_{{\rm NMAD},m}$ and $W_m$.
The weights $w_{\rm acc}, w_{\rm NMAD}$, and $w_{\rm out}$ in Eq~\ref{eq:merit} determine the desired ratio of the merit components.
In this work, we used the respective weight values of $0.4, 0.4$, and $0.2$.
The outlier ratio was given a lower weight because our outlier definition is somewhat arbitrary and it also affects only a small fraction of galaxies.

Based on the calculated merit values (see lower panel of Fig.~\ref{fig:template_grid}), we find that the optimal number of selected templates for bright galaxies is 75, although the differences between other number of chosen templates are not that big for number of templates above 50.
In this work, we were satisfied with only one set of 75 templates as we mostly use the magnitude cut of $r<22$ mag.
But if larger sets of observations should become available in the future, it might be viable to construct different template selections for each brightness range.

\begin{figure}
    \centering
    \includegraphics[width=\linewidth]{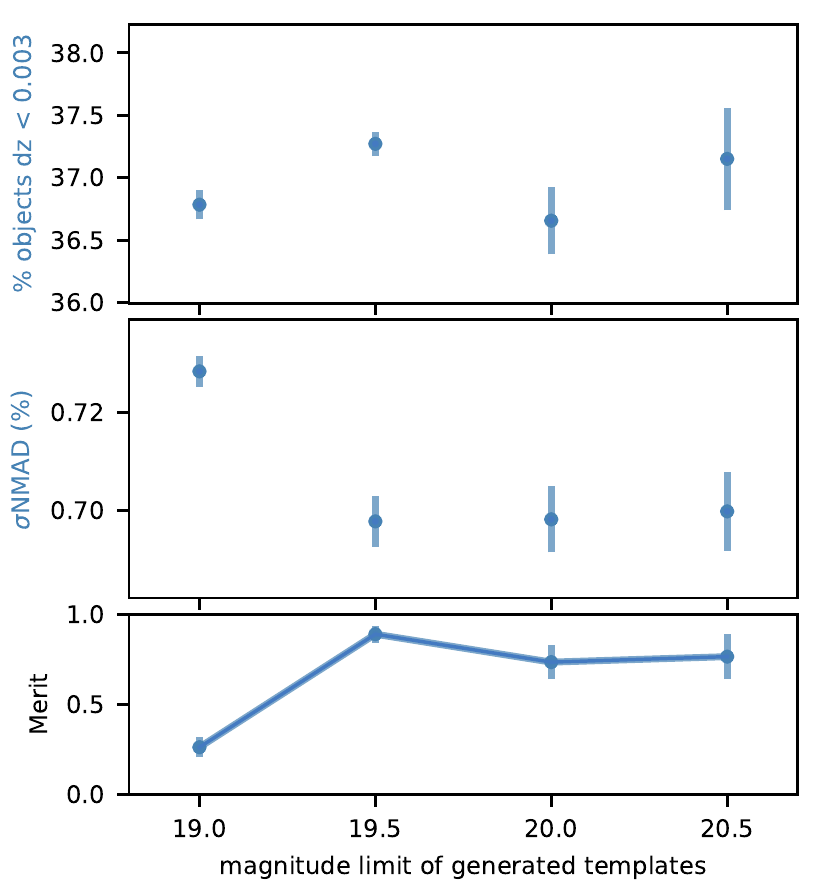}
    \caption{Photometric redshift accuracy depending on the $r$-band magnitude limit of the base templates. The y-axis of each panel is identical to those in Fig.~\ref{fig:template_grid}. Each mark is composed of 10 realisations and each realisation has 100 selected templates.}
    \label{fig:template_grid_mag_lim}
\end{figure}

Another test was carried out to determine whether setting a magnitude limit to the galaxies used for template creation (see Sect.~\ref{sec:template_generation}) has any effect on the final template selection and the resulting redshift estimations.
The reasoning being that by only selecting templates constructed from brighter galaxies, the quality of said templates might be much higher and thus excluding templates constructed from fainter galaxies results in a higher quality template set.

Overall, the magnitude limit does not seem to affect the performance much for template sets of 100 templates (see Fig.~\ref{fig:template_grid_mag_lim}).
The biggest exception being the brightest cut (r $\leq 19$) that considerably decreases the resultant accuracy due to a small number of templates from which the template set can be formed (here, 142). 
Templates based on relatively few brightest galaxies cannot be representative enough for all the observed galaxies. 
If the number of bright sources were larger, such an effect should be significantly reduced. 
However, the possibility that the SEDs of some specific higher redshift sources remain unrepresented in the template set increases with brightness cuts.

\subsubsection{Impact of applying the photo-$z$ priors}
\label{sec:test_prior}

\begin{figure}
    \centering
    \includegraphics[width=\linewidth]{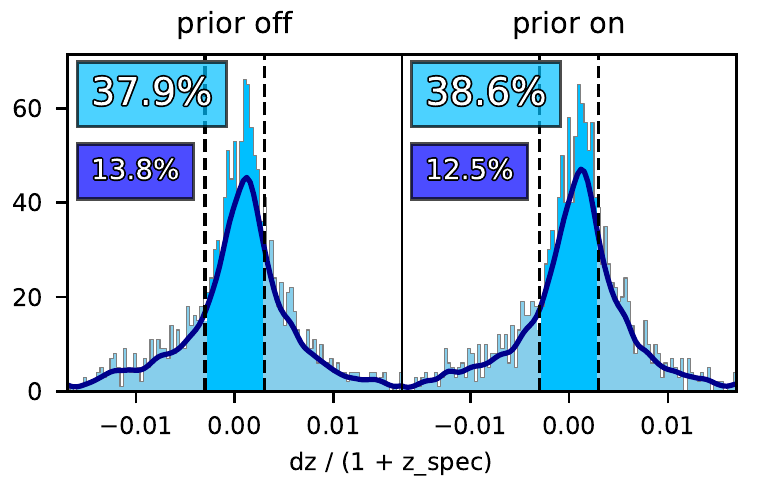}
    \caption{Comparison of the distributions of photometric redshift accuracy with prior turned on and off for galaxies with $r<22$ mag. Galaxies with the accuracy better than the J-PAS target value 0.3\% are marked with darker blue area between the dashed black lines. The percentages in the top left corners show the fraction of objects having the J-PAS target accuracy (upper) and the fraction of outliers (lower).}
    \label{fig:comp_prior_onoff}
\end{figure}

We tested whether the inclusion of a simple prior as described in Sect.~\ref{sec:prior} make a notable impact on the redshift estimation accuracy.
Distribution of the redshift accuracy of the sources with and without prior can be seen in Fig.~\ref{fig:comp_prior_onoff}.
In general, the shape of the accuracy distribution remains roughly the same. 
The minor differences are due to galaxies with the redshift PDF shapes that are most affected by the prior.
The number in the top left corner of each panel shows the ratio of galaxies that achieve the J-PAS target accuracy and the number below that is the fraction of outliers.
As can be seen, there is an improvement of both indicators where the number of outliers is reduced by almost 10 per cent.
Thus, we conclude that applying even a simple prior is an improvement to the overall redshift estimation quality.

\subsubsection{Impact of photometric corrections}
\label{sec:test_corrections}

\begin{figure}
    \centering
    \includegraphics[width=\linewidth]{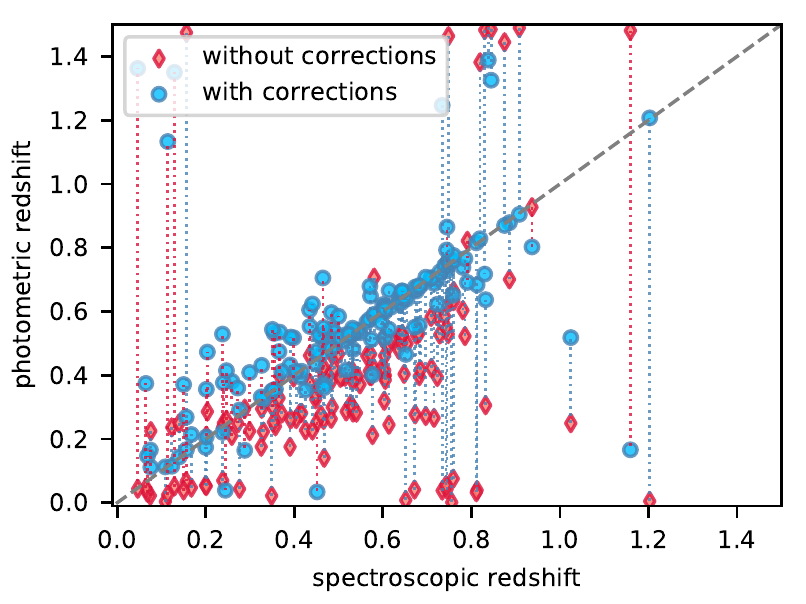}
    \caption{Effect of our photometry corrections to the redshift estimations. Red diamonds are the photo-$z$ values with raw data and blue points are the photo-$z$ values after correcting the photometry. The dotted lines indicate whether the corrections improve (blue line) or worsen (red line) the photo-$z$ estimation.}
    \label{fig:correction_effect}
\end{figure}

The effect of the photometric corrections described in Sect.~\ref{sec:photo_corr} can be seen in Fig.~\ref{fig:correction_effect}.
The red diamonds and blue circles show the {\tt z\_w1d} photo-$z$ estimations before and after applying the photometric correction, respectively.
Only galaxies for which the inclusion of the photometric corrections changed the photo-$z$ estimations by more than 0.1 are presented in the plot.
For these galaxies, we find that the number of improved photometric redshift estimations is $\sim3$ times higher than the number of worsened ones (115 and 36, respectively, out of the total of 1989 galaxies). 
Interestingly, the figure also shows that, on average, the corrections tend to increase the redshift estimation rather than reduce it.
This is most likely due to that linear increase in correction values starting from $\sim8000\:\AA$ (see Fig.~\ref{fig:zp_corrections}).
Only the nearest sources show the opposite effect because our redshift estimations are limited to $z>0$.

However, the galaxies shown in Fig.~\ref{fig:correction_effect} make up only $\sim8\%$ of the whole sample.
For most of the other galaxies, the effect of the photometric correction is much more subtle and for around 75\% of the galaxies in the catalogue, the effect is so negligible that it does not affect their photometric redshift accuracy.

\begin{figure}
    \centering
    \includegraphics[width=\linewidth]{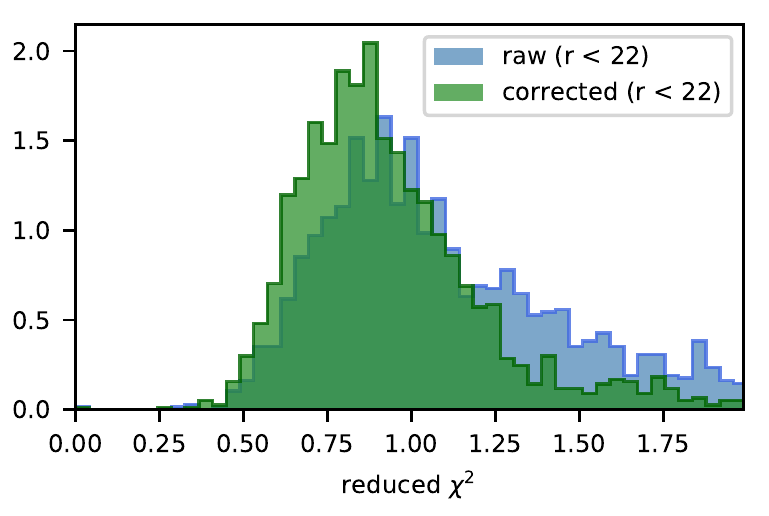}
    \caption{Reduced $\chi^2$ value distributions before (blue) and after (green) applying photometric corrections to the miniJPAS photometry. Compared to the initial catalogue, corrected data has a more condensed reduced $\chi^2$ distribution but the mean value is shifted further away from unity.}
    \label{fig:reduced_chi2}
\end{figure}

We can assess the photometric corrections also from a statistical point of view. 
Assuming that the templates are optimal for the given galaxies, it is expected that the average $\chi^2$ value for the best match template over all passbands (reduced $\chi^2$) remains close to unity.
This means that, on average, the difference between the data and the template is of the same measure as observational errors and, as a result, is affected only by these errors.

The photometric corrections should improve the photometry and lead to better redshift estimations.
This is under the assumption that the templates describe the galaxies somewhat truthfully and that by introducing this correction we will get rid of the systematic offsets embedded in the photometry and bring it closer to the template SEDs.
Fig.~\ref{fig:reduced_chi2} shows the reduced $\chi^2$ values before and after applying the photometric corrections for galaxies in the brighter sub-sample ($r<22$ mag).
The reduced $\chi^2$ value distribution of the initial photometry (blue) is quite broad, with the maximum close to unity and a large wing extending to high values.
By applying the corrections, we get a more narrow distribution. 
The galaxies with the worst photometry improved more than the ones with good photometry.
But this also created a side-effect where the distribution maximum is now shifted to around $\chi^2=0.8$.
While still relatively close to unity, it might also be an indicator of an overestimation of systematic errors either by miniJPAS data processing or as a result of the photometric corrections.
However, we consider the current data set not big enough for far-reaching conclusions and the above deviation from unity big enough for motivating a mechanical reduction of the photometric errors.

\subsection{Comparison with spectroscopic data}
\label{sec:results}

\begin{figure}
    \centering
    \includegraphics[width=\linewidth]{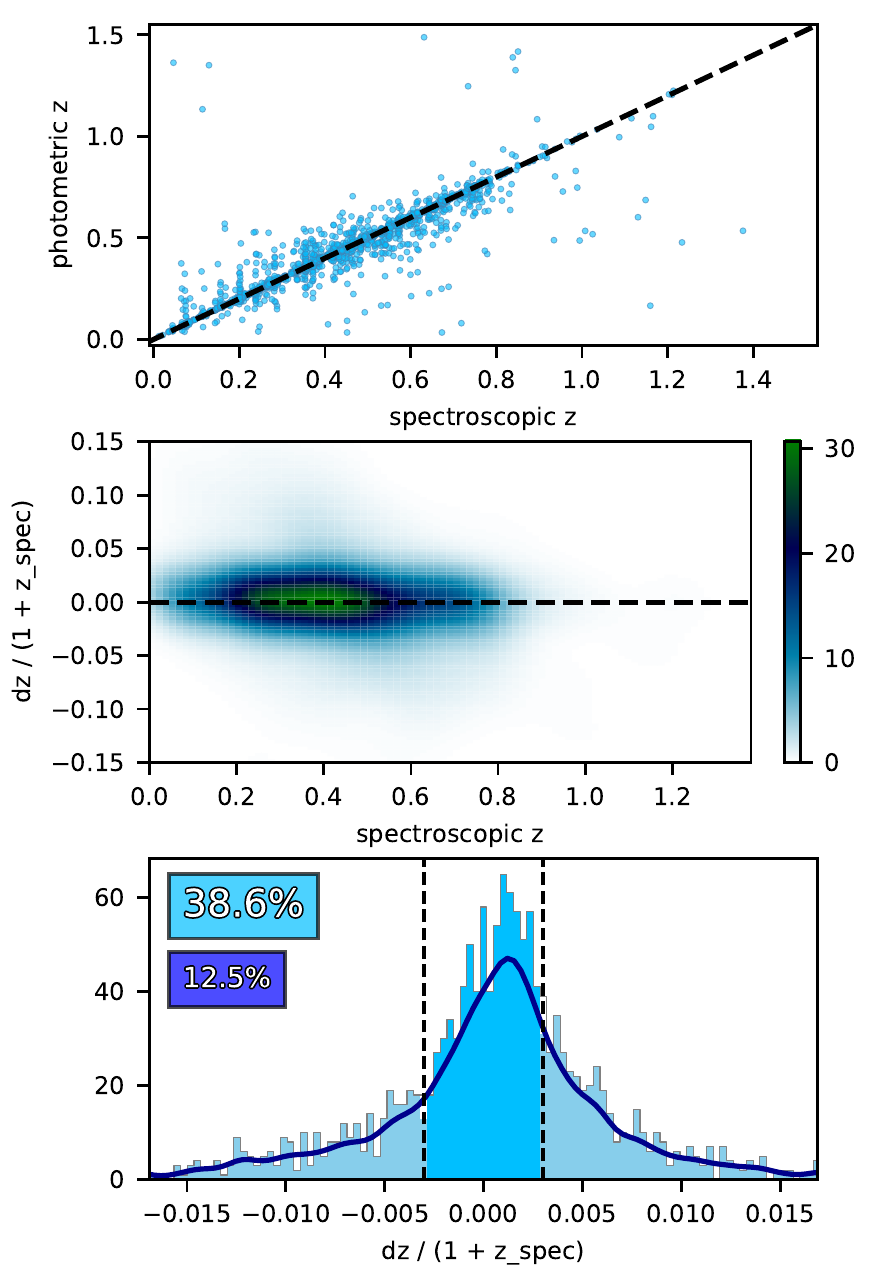}
    \caption{Comparison with spectroscopic redshifts (top panel), surface density of residuals (middle panel), and photometric redshift accuracy distribution (bottom panel) for galaxies $r<22$ mag. Galaxies with a photo-$z$ accuracy better than the J-PAS target accuracy of 0.3\% are marked with darker blue between the dashed lines on the bottom panel. The percentages in the bottom panel show the fraction of objects having the required accuracy and the fraction of outliers in a upper and lower box, respectively.}
    \label{fig:comp_z_spec}
\end{figure}

In order to assess our photo-$z$ estimation accuracy, we have applied TOPz to the miniJPAS catalogue described in Sect.~\ref{sec:mini_jpas} and compared the results with spectroscopic redshifts.
In Fig.~\ref{fig:comp_z_spec}, we give three different types of comparison plots showing the differences between the TOPz {\tt z\_w1d} photo-$z$ results and spectroscopic redshifts for the sample of 1989 galaxies with $r<22$ mag.
The top panel shows a one-to-one comparison of the redshifts.
In general, the miniJPAS-TOPz estimations are in accordance with the spectroscopic values without a major systematic offset trend, although in cases of catastrophic failures, TOPz tends to underestimate redshifts for more distant sources and overestimate them for closer ones. 
The underestimation at higher redshifts may be connected to the above-mentioned lower redshift preference as well as to the colour degeneracy as low redshift red galaxies become similar to high redshift blue galaxies.
The overestimation of lower redshifts has a statistical explanation - it is unlikely to underestimate redshift values that are already close to 0 as negative redshifts are not allowed in TOPz.
The middle panel shows the density of differences between the two as a function of redshift and, on average, no noticeable trend can be seen in the residuals.
The lower panel shows the overall photometric redshift accuracy distribution.
We find that we reach the J-PAS redshift accuracy goal (indicated with dashed vertical lines) for 38.6\% of the galaxies with $r<22$ mag.
We also find that 12.5\% of galaxies have exceedingly wrong redshift estimate, defined as d$z/(1 + z)  > 0.05$, that we consider as outliers.
This is also the current failure rate of the miniJPAS-TOPz combination and, as can be seen below, is similar to the results obtained in \citet{2021A&A...654A.101H}.

One way to increase the reliability of a photometric redshift estimation is the use of the odds parameter.
\citet{2021A&A...654A.101H} showed that by using an odds cut (odds > 0.82) in their catalogue, $\sim 50\%$ of the miniJPAS galaxies reach the J-PAS redshift accuracy goal and only 5\% of the galaxies would be classified as outliers.
We can confirm that when using the same redshift range to calculate the odds values, the TOPz results with odds > 0.82 provides 46\% galaxies that reach the J-PAS redshift accuracy goal and 4\% of galaxies that are classified as outliers.
When using the brighter sub-sample ($r<22$ mag), these percentages would be 48\% and 4\%, respectively.
Therefore, a cut in odds value is, similarly to a cut in brightness, another reliable way to select a sub-sample of galaxies with better photometric redshift estimates.

\begin{figure}
    \centering
    \includegraphics[width=\linewidth]{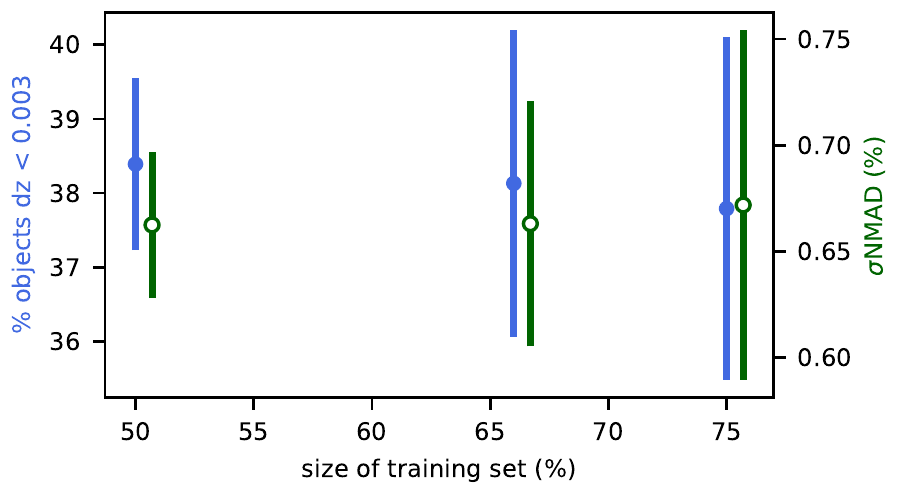}
    \caption{Impact on redshift estimation when two separate subsets are used for training and validation. The x-axis shows the ratio of objects (from to the full catalogue with $r < 22$ mag) that are used for testing. Each marking is comprised of 10 randomly selected subsets and is plotted as the mean value and standard deviation of the distribution of values. The blue points follow the percentage of objects that fit the J-PAS accuracy criteria while the green points follow the achieved median absolute deviation for the normalised distribution ($\sigma$NMAD). The green points are shifted slightly right for better visibility.}
    \label{fig:training_test}
\end{figure}

As discussed in Sect.~\ref{sec:templates}, we are using a set of galaxies from the miniJPAS catalogue for generating our templates and photometric corrections.
The same catalogue is used also to probe the resultant photometric redshift accuracy.
In order to address the possibility that a bias is introduced by using the same catalogue for both training and validating, we have run the TOPz workflow also on two separate subsets of the full catalogue.
The training subset was generated by randomly selecting a fixed number of galaxies from the miniJPAS catalogue and the remainder of the galaxies were left for the validation subset.
The training set was used to generate templates, make a template selection, and calculate photometric corrections.
These templates and photometric corrections were then applied to the validation set to estimate the accuracy of the photometric redshifts.
We varied the corresponding subset sizes to make sure that the potentially small number of galaxies in the training or validation sets does not worsen the statistics.

The impact on the redshift estimation of separating the training and validation subsets can be seen in Fig.~\ref{fig:training_test}.
We ran the TOPz workflow with three different training set sizes and ten randomly selected training sets were generated for each size.
It can be seen that even when using as few as 50\% of the miniJPAS catalogue with $r < 22$ mag for training (i.e. 994 galaxies), we achieve the same fraction of galaxies with $dz < 0.003$ (blue points) than when using the full miniJPAS catalogue for both training and validating.
The same holds true for the $\sigma$NMAD values (green points).
When we increase the training size, the mean values for these characteristics remain almost the same while the variance grows.
This is to be expected, as the number of galaxies in the validation sets keep getting smaller and thus the estimates become more affected by the randomness of the subset selection.
It can also be seen that no bias is introduced when we increase the number of galaxies in the training subset.
Therefore, we can be confident that using the same catalogue for both training and validating the photo-z estimates does not bias the final results.

\begin{figure}
    \centering
    \includegraphics[width=\linewidth]{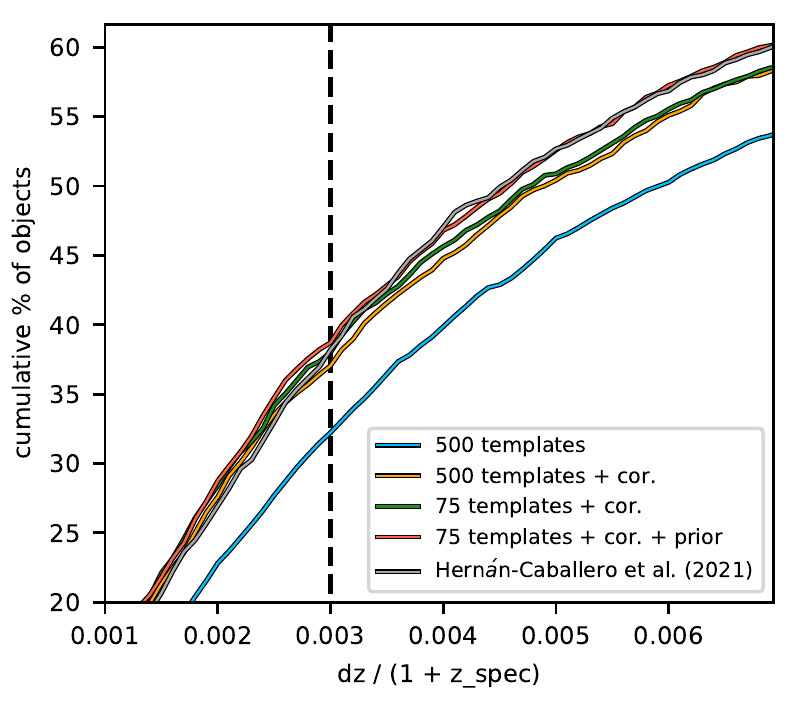}
    \caption{Fraction of galaxies satisfying accuracy limits. Different limits are shown on the x-axis with the J-PAS target accuracy of 0.3\% indicated with a dashed vertical line. Colours show different TOPz runs using various setups described in this paper and grey corresponds to the LePhare photo-$z$ estimations presented in the miniJPAS database. This plot shows how correcting the observations and adding photo-$z$ priors improve the results significantly. Galaxies with $r<22$ mag and TOPz output {\tt z\_w1d} were used.}
    \label{fig:comp_diff_improvement}
\end{figure}

Figure~\ref{fig:comp_diff_improvement} shows how the improvements described in Sect.~\ref{sec:prior_template} affect photometric redshift estimations. 
The results improve progressively when we use the full base template set (blue), enable photometric corrections (orange), apply the reduced template selection (green), and finally enable priors along with the previous steps (red).
The grey line shows the photo-$z$ estimations from miniJPAS database and the dashed vertical line marks the target accuracy of J-PAS redshifts.
Every added step increases the number of galaxies achieving J-PAS photo-$z$ target accuracy with the biggest improvement coming from implementing the photometric corrections (see Sect.~\ref{sec:photo_corr}).
Overall, TOPz results are on par with previous miniJPAS photo-$z$ estimations showing the accuracy that template-based photometric redshift methods can reach given the quality of the miniJPAS photometry.

\begin{figure}
    \centering
    \includegraphics[width=\linewidth]{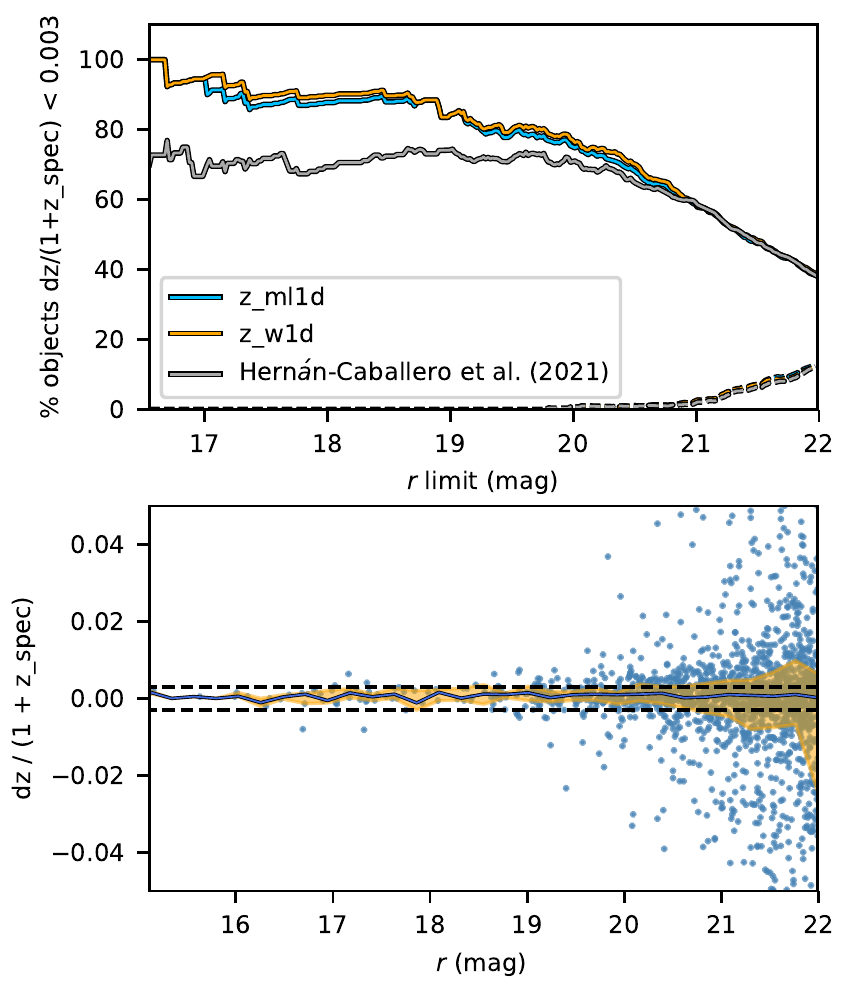}
    \caption{TOPz results based on the brightness of galaxies.
    Upper panel: fraction of galaxies within the J-PAS target accuracy (solid lines) and outliers (dashed lines) depending on the limiting magnitude of the sample. Colours show {\tt z\_ml1d} (blue) and {\tt z\_w1d} (orange) redshift estimators from TOPz and results from miniJPAS database (grey). 
    Lower panel: photometric redshift accuracy for individual galaxies alongside their apparent brightness (blue points). The blue line marks the binned median value, the shaded area is where 50\% of the binned data lies and the dashed black lines mark the 0.3\% accuracy limit.}
    \label{fig:mag_vs_ratio}
\end{figure}

All of the previous results and plots were shown for sources with $r<22$ mag.
From the observational perspective, the brighter the galaxies are, the less noisy the observations become and the easier it is to estimate photometric redshifts.
The ratio of TOPz photo-$z$ estimations reaching the \mbox{J-PAS} target accuracy depending on the depth of the observations can be seen in the upper panel of Fig.~\ref{fig:mag_vs_ratio}.
As expected, the photo-$z$ accuracy falls when taking more fainter galaxies into account.
As discussed in Sect.~\ref{sec:test_templates}, fainter galaxies would benefit from a smaller set of templates.

Different TOPz redshift estimators (maximum likelihood {\tt z\_ml1d} and weighted {\tt z\_w1d}) yield a similar dependency on source magnitude (see Fig.~\ref{fig:mag_vs_ratio}).
At the bright end, the object density is low and the scatter in estimators is caused by statistical errors.
The weighted estimator is slightly better up to 21 mag and becomes equal to other estimators when taking fainter galaxies into account.
The dashed lines in the upper panel of Fig.~\ref{fig:mag_vs_ratio} show the corresponding outlier fraction.
As expected, this fraction increases for fainter sources and no real variation in different photo-$z$ estimators can be seen.
The accuracy of \citet{2021A&A...654A.101H} photo-$z$ results (presented with grey) is somewhat worse than TOPz for brighter galaxies whereas the accuracy becomes equal for the full catalogue.
The differences between \citet{2021A&A...654A.101H} and TOPz results on the brighter end may be caused by different optimisation of templates or other choices in the configuration. 
However, a  detailed analysis of these differences is meaningless given the small size of the galaxy sample and a more thorough assessment have to wait until more data from the full J-PAS survey become available.

The lower panel of Fig.~\ref{fig:mag_vs_ratio} shows the photo-$z$ accuracy for individual galaxies.
We note that some outliers are left out of the bounds of the plot as otherwise the more interesting central region would become too tiny for visualisation.
Galaxies with apparent magnitudes up to $\sim19$ are mostly within the J-PAS accuracy limit, shown with the dashed black line.
The accuracy becomes worse at fainter magnitudes and from above $\sim21$ mag, less than half of the galaxies remain within the J-PAS target accuracy.

\subsection{Understanding the PDF}
\label{sec:PDF_result}

Although a single redshift value is usually the most preferred output of a redshift catalogue, the underlying redshift PDFs contain more useful information.
As one of the TOPz outputs is the full redshift PDF of every template-galaxy pair, we can conduct some statistical tests on the whole catalogue to determine how well-behaved our redshift PDFs are in terms of statistics.

In \citet{2016arXiv160808016P}, the probability integral transform (PIT) is used as a diagnostic tool to check the calibration and the sharpness of the generated predictive distributions.
The PIT is easily validated visually as a histogram of the cumulative probabilities at the value of the spectroscopic redshift.
Only if the PIT histogram bins are distributed uniformly, are the PDFs well calibrated.
Whereas, if the histogram is u-shaped, the PDFs are overconfident and in the opposite case, the PDFs are underconfident.

\citet{2016arXiv160808016P} also introduced a continuous ranked probability score (CRPS) to measure the individual performances of the PDFs.
CRPS expresses the distance between the PDF and the spectroscopic redshift value and is used to measure how precise is the PDF shape in relation to the true value.
An average CRPS value is used to estimate the overall performance of a redshift estimation method on a given galaxy set, with a lower value indicating better PDFs.

Finally, a test to measure the fraction of galaxies ($\hat{F}$(c)) in which the spectroscopic redshift falls within a given confidence interval (CI) of the PDF is used to determine how well calibrated the PDFs are.
\citet{2016MNRAS.457.4005W} define the CI as the area under the PDF that reaches the probability threshold calculated at the spectroscopic redshift.
Likewise to the case with PIT, if the PDFs are calibrated well then we can expect 10\% of galaxies to fall within 10\% CI, 20\% within 20\% CI, etc.
Therefore, when the cumulative fraction of galaxies $\hat{F}$(c) follows the CI values, the PDFs are well calibrated whereas when the $\hat{F}$(c) is below (above) the CI relation, the PDFs are deemed over- (under-) confident.

\begin{figure}
    \centering
    \includegraphics[width=\linewidth]{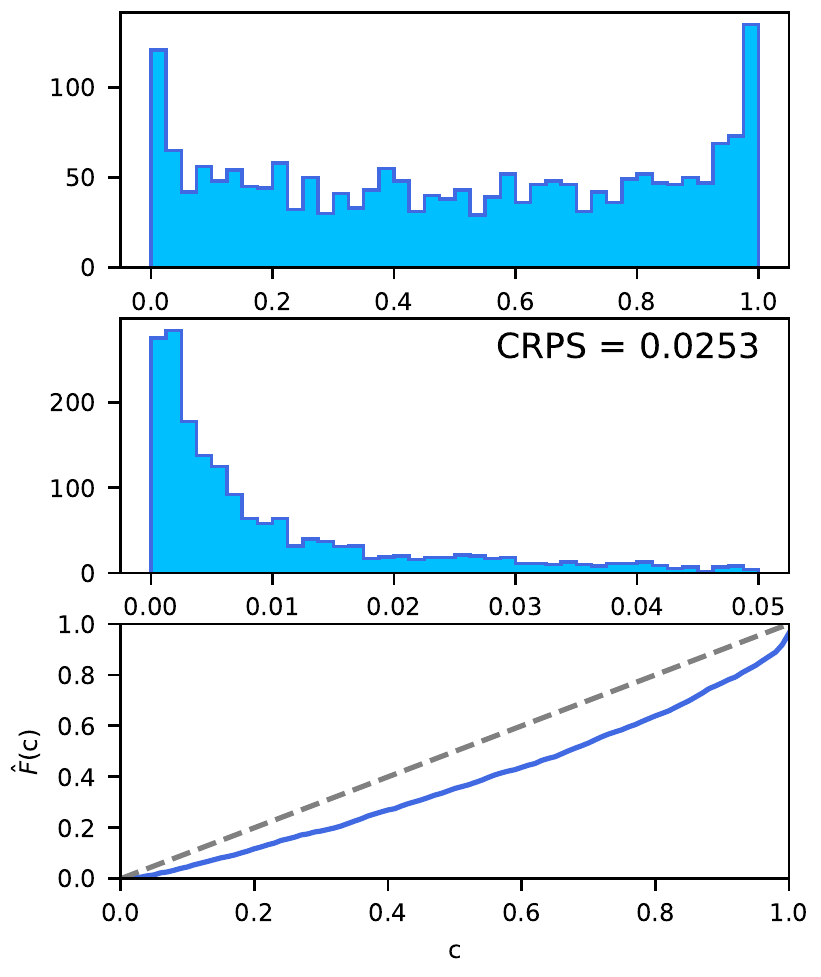}
    \caption{PIT (upper panel), CRPS (middle panel), and $\hat{F}$(c) (lower panel) tests for the $r<22$ mag catalogue. The average CRPS value is given in the middle panel. The dashed grey line on the lower panel shows the expected $\hat{F}$(c) - c relation.}
    \label{fig:error1_pdf}
\end{figure}

\begin{figure}
    \centering
    \includegraphics[width=\linewidth]{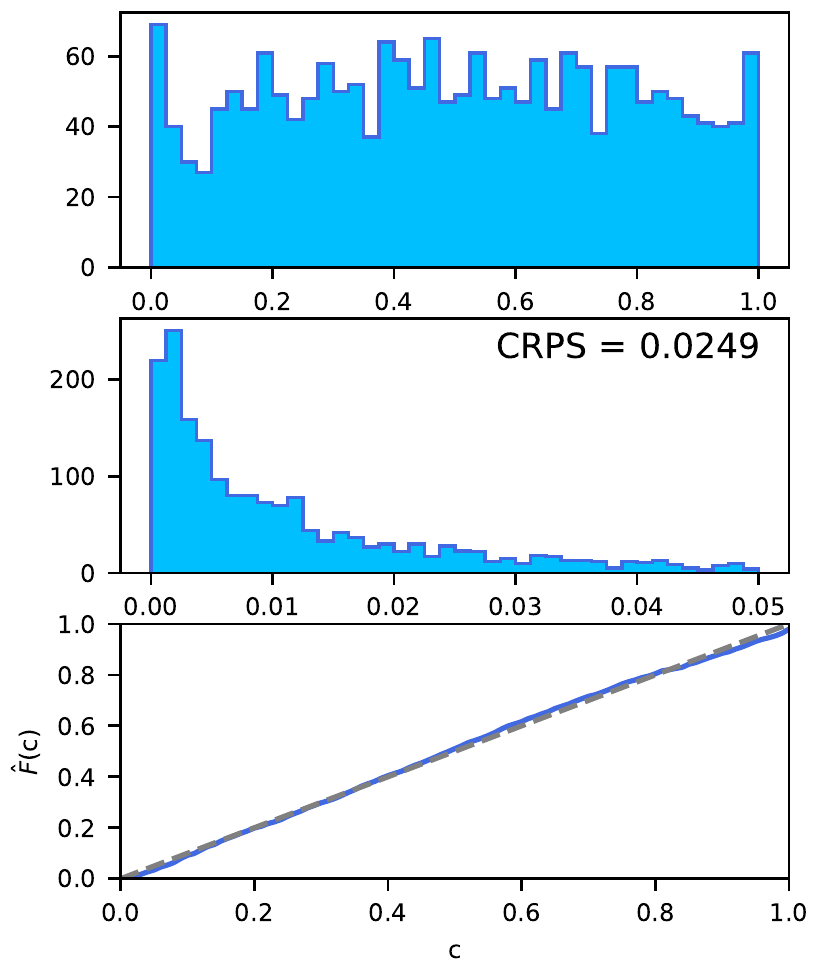}
    \caption{Same panels as in Fig.~\ref{fig:error1_pdf}, but with the uncertainties of the $r<22$ mag catalogue increased by 1.3 times.}
    \label{fig:error13_pdf}
\end{figure}

We have analysed the PDFs of the $r<22$ mag catalogue galaxies using three above-mentioned statistics and the results can be seen in Fig.~\ref{fig:error1_pdf}.
Both the PIT as well as the $\hat{F}$(c) - c relation show that the PDFs of our miniJPAS catalogue are somewhat overconfident.
This means that the PDFs are too narrow and do not correspond well to the true redshift. 
This is usually a problem resulting from underestimating the observational uncertainties.

In Fig.~\ref{fig:error13_pdf}, we conducted the same analysis but this time we increased the photometric uncertainties by 1.3 times.
After increasing the uncertainties, the overconfidence in the PDFs are eliminated and the calibration is much more uniform.
Also the CRPS score is lower than it was for the original catalogue, meaning that the PDFs are, on average, better defined.
Therefore, we can conclude that the photometric uncertainties of the miniJPAS catalogue are under-valuated.
Although, the PDFs are better described with increased uncertainties, the resulting redshift estimates are $\sim2\%$ worse than what we achieved in Sect.~\ref{sec:results}.
In order to to keep the best redshift estimates while having better calibrated PDFs, one option is to apply a contrast correction to the original PDFs as was done by \citet{2021A&A...654A.101H}.

\begin{figure}
    \centering
    \includegraphics[width=\linewidth]{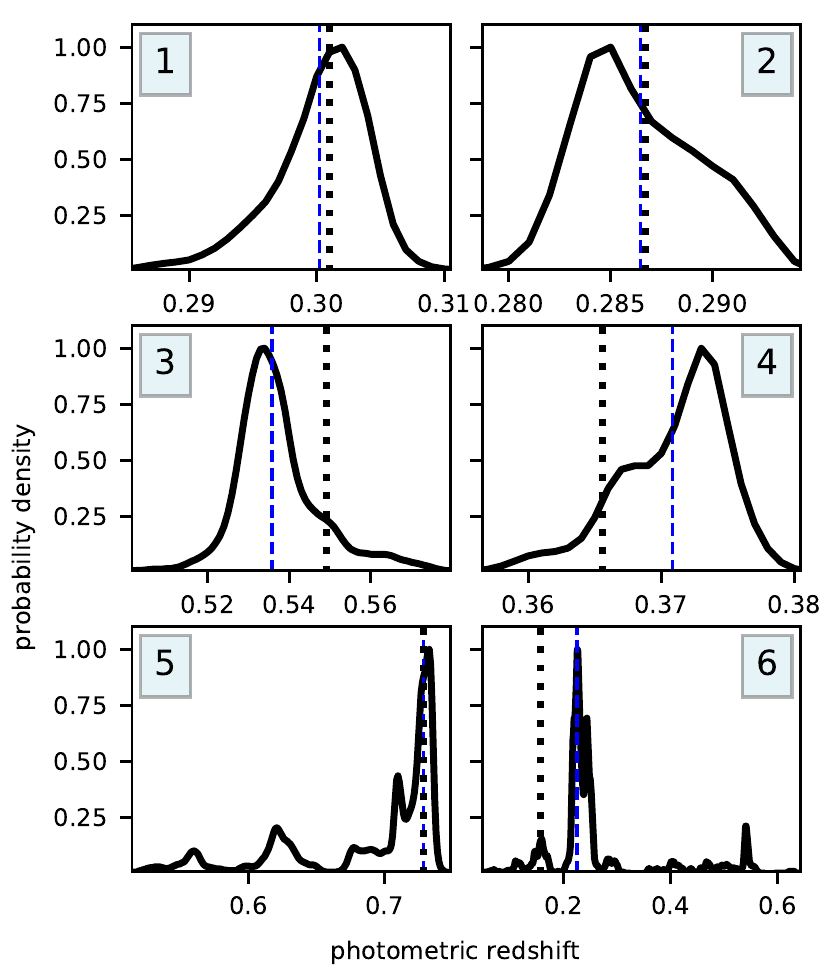}
    \caption{Examples of different redshift PDF shapes. Spectroscopic redshift values are marked with vertical black dotted lines and TOPz redshift estimations ({\tt z\_w1d}) with dashed blue lines.}
    \label{fig:random_posteriors}
\end{figure}

The redshift PDFs of different galaxies can be quite diverse as the accuracy of the observed photometry will cause various interactions with the templates.
Some examples of the PDF shapes are given in Fig.~\ref{fig:random_posteriors} with the black dotted line marking the spectroscopic redshift and the dashed blue line the TOPz weighted ({\tt z\_w1d}) redshift estimation.
With very prominent emission lines, the PDF shapes would have a single symmetrically narrow peak.
In more general cases, some PDF shapes are relatively simple with one maximum and slightly lopsided wings (panel 1 and 2) whereas other shapes can be more complex with multiple peaks and valleys.
The latter cases stem from photometric uncertainties as well as from the types of templates dominating in the PDF construction.
There are tougher cases where the templates get quite similar $\chi^2$ values for a wider area around the true redshift while different types of galaxy templates have their maximum set at different redshifts.
When we marginalise them into the PDF, we are left with a shape where the spectroscopic redshift is at the secondary maximum on the side of a more dominant peak (panel 3 and 4).
This is the main motivation to prefer the weighted average around the maximum ({\tt z\_w1d}) instead of the peak location, as the former takes into account also the lopsidedness of the PDF without losing much precision in other cases. 
However, it does not help much in the worst cases.
For more complex shapes (panel 5 and 6), the difference in templates is more distinct but the spectroscopic redshift value is not always around the maximum and there is no universal way to detect and solve these cases.

\begin{table}
    \centering
    \caption{TOPz estimator performance for $r<22$ mag galaxies with multi-peaked PDFs. }
    \label{tab:topz_estimator_multipeak}
    
    \begin{tabular}{cccc}
        \hline\hline
        TOPz estimator & d$z<0.003$\tablefootmark{a} & d$z>0.05$\tablefootmark{b} & min(d$z_{\rm{phot}}$)\tablefootmark{c}\\
        \hline
        \rule{0pt}{2ex}
        {\tt z\_ml1d} & 15.7\% & 27.1\% & 27.1\%\\
        {\tt z\_w1d} & 17.9\% & 25.7\% & 35.0\%\\
        {\tt z\_ml2d} & 6.4\% & 42.1\% & 17.1\%\\
        {\tt z\_w2d} & 7.9\% & 41.4\% & 20.7\%\\
        
        \hline
        
    \end{tabular}
    
    \tablefoot{
    \tablefoottext{a}{Percentage of galaxies that reach the J-PAS redshift accuracy goal using this TOPz estimator.}\\
    \tablefoottext{b}{Percentage of galaxies that we consider outliers using this TOPz estimator.}\\
    \tablefoottext{c}{Percentage of objects for which this TOPz estimator is closest to the spectroscopic redshift.}
    }

\end{table}

Specifically, when comparing {\tt z\_w1d} and {\tt z\_w2d} redshift estimations, the two values coincide for 93\% of the galaxies in our test catalogue.
This means that 7\% of the galaxies are multi-peaked (similar to panel 5 and 6 in Fig.~\ref{fig:random_posteriors}).
In general, there are no excellent redshift estimators for these cases.
We find that {\tt z\_w1d} performs a little bit better having 17.9\% of galaxies with multi-peaked PDFs fall into the J-PAS accuracy limit compared to the 7.9\% with {\tt z\_w2d}.
When compared directly, for 62.8\% of these galaxies, the {\tt z\_w1d} value is closer to the spectroscopic redshift than the {\tt z\_w2d} value.
When also taking into account the {\tt z\_ml1d} and {\tt z\_ml2d} estimators, {\tt z\_w1d} value is closest to the spectroscopic redshift for 35.0\% of them and also remains the most accurate TOPz estimator within the J-PAS accuracy limit.
Table~\ref{tab:topz_estimator_multipeak} summarises the performance of the four TOPz redshift estimators for galaxies with multi-peaked PDFs.
Although, on average, {\tt z\_w1d} performed the best out of the current TOPz redshift estimators, there are still specific cases where other estimators are better thus a more robust estimator selection could be made with more available data.

\section{Conclusion and discussion}
\label{sec:conclusion}

We have developed a software package TOPz for template-based photometric redshift estimation, specifically designed for the forthcoming multi-filter narrow-passband J-PAS survey data. 
We applied TOPz on the precursor observations to the J-PAS, the miniJPAS, and compared the redshift estimations to the spectroscopic redshifts from the DEEP2, DEEP3, and SDSS observations.
The spectral templates of galaxies were generated on the basis of the actual miniJPAS sources using the synthetic spectrum generation software \textit{CIGALE}. 
We showed that reducing the number of base templates improves the photo-$z$ accuracy and that the number of optimal templates depends on the brightness cut of a given galaxy sub-sample.
We also showed that applying the photo-$z$ priors as well as correcting the photometry using the template library can further improve the accuracy.
We find that, using the redshift estimations from TOPz, the overall number of galaxies satisfying the J-PAS redshift accuracy goal is on par with previous results.
However, as both the observed area of miniJPAS as well as the number of observed galaxies with known spectroscopic redshifts are quite small, the statistical uncertainties reduce the possibilities for a more detailed assessment of the results and the verification of specific choices made in the workflow.
With future observations, the results could be further improved by using better-calibrated templates, priors, and other input information.

In particular, an important J-PAS-specific improvement in TOPz is the possibility to take into account the dependence of the central wavelengths of the filter transmission bands on the incident angle of the light path \citep{2014arXiv1403.5237B}. 
This dependence means that each galaxy in each filter is observed through a slightly different passband depending on the physical position of the galaxy's image on the surface of the filter during an exposure. 
As only the average passband of each filter of J-PAS is provided at the moment, taking such passband shifts into account would improve the match between synthetic and J-PAS photometry. 
Currently, this information is lost during the combination of different exposures into \lq tiles\rq, but can be taken into account with the main J-PAS data releases.
For more detailed analysis on how we expect the results to improve, see Appendix.~\ref{sec:CW}.

A number of required and optional inputs (see Sect.~\ref{sec:topz_overview}) affect the resulting photometric redshift estimations with TOPz.
Unfortunately, the effects of these inputs are often degenerate and a set of inputs that improve the results separately might not do so when combined.
TOPz is most strongly affected by the input templates, observational data quality, the accuracy of the uncertainty estimates, and photo-$z$ priors.

Out of the listed factors, we only had considerable control over the templates. 
We tested different approaches for optimising the set of templates. 
The optimal base template set depends on the overall configuration of the workflow (with priors, without priors, with priors and added corrections, and so on).
In addition, inputs that produce good results with one observational data set may not be the best for others.
Most notably, template set tuned for brighter sources is not optimal for fainter, noise-dominated sources.
The same may apply for the targeted redshift range; however, the current data did not allow us to investigate such dependencies.
Similarly, we can expect that more sophisticated prior, which considers the expected abundance of a given spectral type at a given redshift, can further improve the accuracy of redshifts. 
The necessary observational information can come from the same data being analysed, but the number of sources has to be considerably larger.
For best results, it is necessary to adjust the inputs separately for each specific set of data and also the science goal.

On a final note, the maximum as well as the overall distribution of the redshift accuracy histogram are slightly shifted towards positive values (see Fig.~\ref{fig:comp_z_spec}).
This means that, on average, the photometric redshifts are overestimated. 
The same holds true for the photometric redshifts available in the miniJPAS database, described in more detail in \citet{2021A&A...654A.101H}.
The shift seems to be independent on the method used to extract the photometric redshifts and might thus be caused by some hidden aspect of the miniJPAS data.
Regardless of the underlying reason, a simple systematic shift applied to all photo-$z$ estimations would improve the overall accuracy.
A more detailed analysis of this systematic effect must wait for larger data sets from future observations.

This work was meant as a precursor to the photometric redshift estimations for the full J-PAS observations as well as any other large-scale photometric redshift survey.
As the number of galaxies in miniJPAS that also have reliable spectroscopic redshift measurements is small, we expect that some of the problems and limitations we faced will not be present in the full survey.
The same should hold true for the systematic observational effects that miniJPAS data exhibited as the telescope setup will be different in several aspects, including the camera, filter tray system, and the field of view. 
Thus, while we showed that the combination of TOPz and miniJPAS data generally already fulfils the expectations for J-PAS redshift accuracy, we expect some further improvement once the actual full J-PAS data start to become available.

\begin{acknowledgements}
      We thank the anonymous referee for useful comments and suggestions, which have enabled us to improve the paper. Part of this work was supported by institutional research funding IUT40-2, JPUT907, PSG700 and PRG1006 of the Estonian Ministry of Education and Research. We acknowledge the support by the Centre of Excellence 'Dark side of the Universe' (TK133) financed by the European Union through the European Regional Development Fund.
      J.C.M. acknowledges partial support from the Spanish Ministry of Science, Innovation and Universities (MCIU/AEI/FEDER, UE) through the grant PGC2018-097585-B-C22.
      L.A.D.G. and R.M.G.D. acknowledge financial support from the State Agency for Research of the Spanish MCIU through the 'Center of Excellence Severo Ochoa' award to the Instituto de Astrof\'\i sica de Andaluc\'\i a (SEV-2017-0709) and to the projects PID2019-109067-GB100 and AYA2016-77846-P.
      J.A.F.O. acknowledges the financial support from the Spanish Ministry of Science and Innovation and the European Union -- NextGenerationEU through the Recovery and Resilience Facility project ICTS-MRR-2021-03-CEFCA.
      This work is based on observations made with the JST/T250 telescope and JPCam at the Observatorio Astrofísico de Javalambre (OAJ), in Teruel, owned, managed, and operated by the Centro de Estudios de Física del Cosmos de Aragón (CEFCA). 
      We acknowledge the OAJ Data Processing and Archiving Unit (UPAD) for reducing and calibrating the OAJ data used in this work.
      Funding for the J-PAS Project has been provided by the Governments of Spain and Aragón through the Fondo de Inversión de Teruel, European FEDER funding and the Spanish Ministry of Science, Innovation and Universities, and by the Brazilian agencies FINEP, FAPESP, FAPERJ and by the National Observatory of Brazil with additional funding also provided by the Tartu Observatory and by the J-PAS Chinese Astronomical Consortium.
      This paper has gone through internal review by the J-PAS collaboration.
      
\end{acknowledgements}

\bibliographystyle{aa} 
\bibliography{mybib} 

\begin{appendix}

\section{Assessing the effect of miniJPAS filter passbands on the redshift estimation.}
\label{sec:CW}

\begin{figure*}[!ht]
    \centering
    \includegraphics[width=17cm]{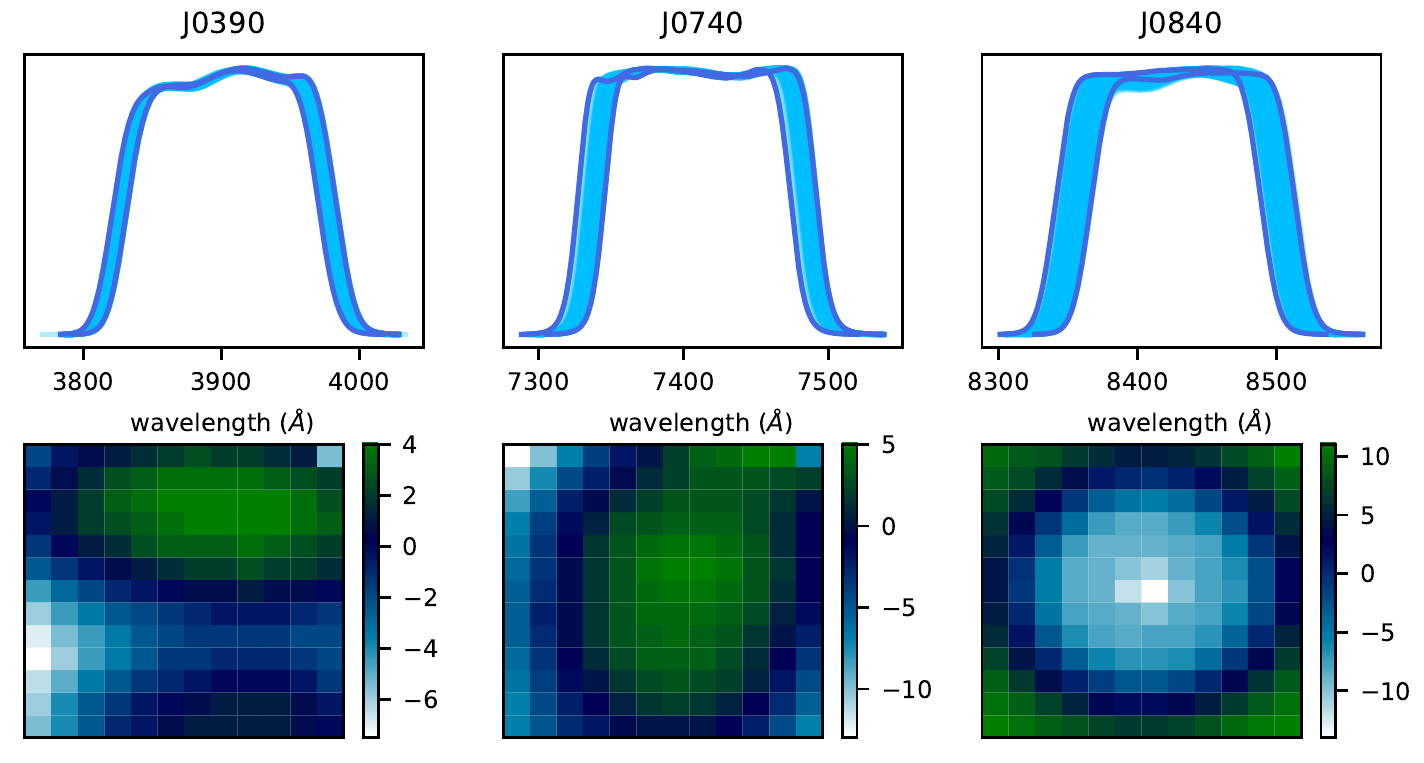}
    \caption{Measured passbands and their overall shape on the physical filter for three different J-PAS filter. All the passbands (top panel) on a single filter were measured in a 12 by 13 grid (bottom panel). Darker blue lines on the top panels show the two most opposed transmission curves of that filter. The colourbar on the bottom panels indicates each passband's central wavelength difference from the mean passband.}
    \label{fig:cw_examples}
\end{figure*}

In order to estimate photometric redshifts, one required input to the photo-$z$ code is the transmission curve of the optical system through which the object is observed.
An important component in this total transmission curve is the filter transmission which can differ depending on the incident angle of the light.
This change in the transmission has been measured for every miniJPAS filter on a 13 by 12 grid covering the full physical surfaces of the filters (see Fig.~\ref{fig:cw_examples}). 
The top panels on Fig.~\ref{fig:cw_examples} give an indication how much the transmission can change due to the incident angle of the light and the bottom panels show the central wavelength difference when compared to the average passband in that filter.
As can be seen on the bottom panel, the change in the central wavelength of the transmission curves do vary smoothly across a single filter but the overall shape differs a lot from filter to filter. 
Therefore, it is hard to predict what the total effect of these changes are when all of the filters are used together in the photo-$z$ codes.

Right now, it is not possible to determine which part of the filter was used in the actual miniJPAS observations and therefore no tests can be done using the observed catalogue.
We tried to estimate the effect of the transmission curve differences on the redshift estimation by simulating the observations as if observed through different areas on this 13 by 12 grid.
For this, we used the remaining templates that were left after the template selection process described in Sec.~\ref{sec:template_selection} and constructed redshifted synthetic photometry catalogues for seven fixed areas from the 13 by 12 grid.
Each catalogue consists of synthetic observations from all of the remaining templates in all of the miniJPAS filters and as if observed through the same passband that is measured through the selected area on every filter.
As the templates themselves were calculated using actual miniJPAS observations, we could simulate synthetic observational uncertainties by applying a gaussian error to the synthetic photometry using the actual photometric errors measured in each corresponding filter.
Then, the synthetic catalogues were run through TOPz workflow to compare how the resulting photo-$z$ accuracy changed depending on the passbands applied to the templates.
To get an idea of the effect of the applied observational uncertainties, we ran each simulation three times.

\begin{figure}
    \centering
    \includegraphics[width=\linewidth]{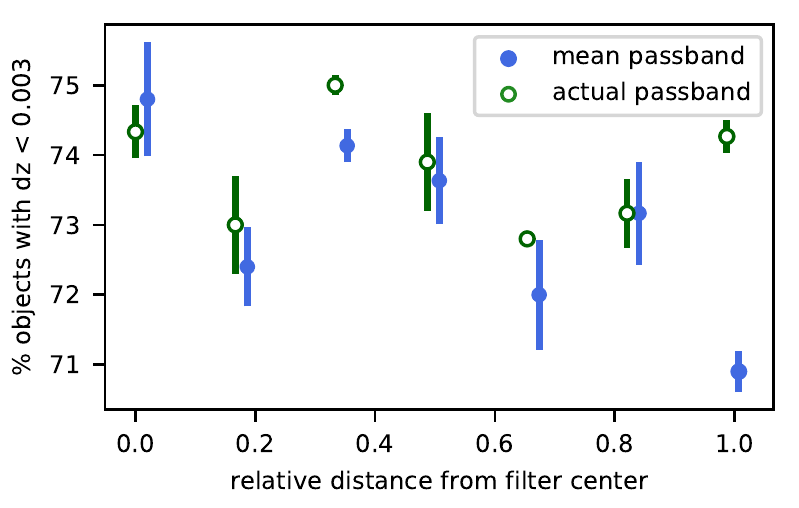}
    \caption{Effect of the variable filter passband on the redshift estimation accuracy. Blue markers note the accuracy when the average passband is used and green markers when the specific passband is used. The relative distance shows the physical distance from the centre of the filter towards the lower-left corner. Blue markers are slightly separated for visual clarity.}
    \label{fig:CW_pixel_vs_dz}
\end{figure}

Figure~\ref{fig:CW_pixel_vs_dz} shows the accuracy of photometric redshift estimations through TOPz depending on whether the input passbands are given as average passbands or as filter area specific passbands.
The $x$-axis shows the relative distance from the filter centre and all of the selected filter areas are directed towards the lower-left corner of the filter.
Every marking at different distance value consists of simulations where a synthetic photometry catalogue is constructed as if observed through that specific area on the filter and the colour indicate the passband that is given as an input during a single TOPz run.
Blue markings are shifted for visual clarity and should be counted as coinciding with the nearby green markings.
Accuracy errors on Fig.~\ref{fig:CW_pixel_vs_dz} are calculated as a standard deviation of the three simulation runs where the photometric errors in each synthetic catalogue were applied in a random manner.
Keep in mind that the filter transmission measurements are actually made on a grid so the relative distance from the filter centre in Fig.~\ref{fig:CW_pixel_vs_dz} notes a single rectangular area on the filter.
As the shapes of the central wavelengths differences are not radially symmetric, the results should differ if moving into any other direction from the centre.
Nevertheless, it can be seen that in almost every filter area, the accuracy is better when a specific passband for that area is used as opposed to the mean passband that we used throughout the analysis in this paper.
Therefore, in order to improve the photometric redshift estimations in the full J-PAS catalogue, it is necessary to map the CCD coordinates of the J-PAS observations to the actual physical coordinates on the filter and then use this info to construct a different passband tailor-made for each observed object.

\end{appendix}

\end{document}